\documentclass[aps,prd,preprint,tightening,showpacs]{revtex4-2}
\usepackage{amsmath,amssymb,amsthm,graphicx}
\usepackage[mathscr]{eucal}
\usepackage[colorlinks,linkcolor=blue,
anchorcolor=blue,citecolor=green]{hyperref}
\usepackage[dvipsnames,usenames]{color}
\usepackage{xcolor}

\newcommand{\be}{\begin{equation}}
\newcommand{\ee}{\end{equation}}
\newcommand{\ba}{\begin{eqnarray}}
\newcommand{\ea}{\end{eqnarray}}
\usepackage{epsf,epsfig,graphics}
\usepackage{verbatim,color,ulem}
\begin{document}
\title{Inspiral and Plunging Orbits in Kerr-Newman Spacetimes}
\author{Yu-Chung Ko}
\author{Da-Shin Lee}
\email{dslee@gms.ndhu.edu.tw}
\author{Chi-Yong Lin}
\email{lcyong@gms.ndhu.edu.tw}
\affiliation{
Department of Physics, National Dong Hwa University, Hualien, Taiwan, Republic of China}
\date{\today}

\begin{abstract}
We present the analytical solutions for the trajectories of particles that spiral and plunge inward the event horizon along the timelike geodesics following general non-equatorial paths within Kerr-Newman spacetimes.
Our studies encompass both bound and unbound motions.
%
The solutions can be written in terms of the elliptical integrals and the Jacobian elliptic functions of manifestly real functions of the Mino time.
They can respectively reduce to the Kerr,  Reissner-Nordstr$\ddot{o}$m, and Schwarzschild black holes in certain limits of the spin and charge of the black holes, and can be compared with the known ones restricted in equatorial motion.
%
{These explicit solutions may have some implications for the gravitational wave emission from extreme mass-ratio inspirals.}

\end{abstract}

\pacs{04.70.-s, 04.70.Bw, 04.80.Cc}

\maketitle
\newpage
\section{Introduction}
%
Recent detections of gravitational waves emitted from the merger of binary systems have confirmed Einstein's century-old prediction as a consequence of general relativity \cite{LIGOS:2016,LIGOS:2018,LIGOS:2020}.
The capture of the spectacular images of supermassive black holes M87* at the center of M87 galaxy \cite{M87:2019_1}
and Sgr A* at the center of our galaxy \cite{SgrA:2022_1}
leads to another scientific achievement of a direct evidence of the existence of  black holes.
The black hole is one of the mysterious stellar objects, and it is the solution derived from the Einstein's field equations \cite{MIS,CHAS}.
In astrophysics, extreme mass-ratio inspirals (EMRIs), which consist of a stellar mass object orbiting around a massive black hole, have recently received  considerable attention.
The goal is to analyze gravitational wave signals to accurately test the predictions of general relativity in its strong regime.
Gravitational wave signals generated through EMRIs, which are  key sources of low frequency  gravitational waves and to be observed in the planned space-based Laser Interferometer Space Antenna (LISA), provides an opportunity to measure various fascinating properties of supermassive black holes \cite{eLISA,Bar,LISAConsortiumWaveformWorkingGroup_2023,Bab,Koc}.

The present work is motivated by EMRIs, which can be approximated as a light body travels along the geodesic of the background spacetime of a massive black hole.
In particular, the recent studies in \cite{mummery2022inspirals,Mummery_2023} have been devoted to inspirals of the particle on the equatorial plane asymptotically from the innermost stable circular orbits (ISCO) of Kerr black holes.
They also derive a simple expression for the equatorial radial flow from the ISCO relevant to the dynamics of the accretion disk.
%
{These exact solutions may have applications to the generated gravitational waveforms arising from EMRIs \textcolor{red}{ \cite{LISAConsortiumWaveformWorkingGroup_2023}}} as well as constructing theories of black hole accretion \cite{FAB,PAG,REY,SCH,JAI}.
Additionally, the work by \cite{DYS,DYS1} broadens the investigation of equatorial motion to encompass generic non-equatorial orbits in Kerr black holes.
%
%
In the family of the Kerr black holes
 due to the underlying spacetime symmetry the dynamics possesses two conserved quantities, the energy {$E_{m}$} and the azimuthal angular momentum {$L_{m}$} of the particle.
Nevertheless, the existence of the third conserved quantity, discovered in the sixties and known nowadays as Carter constant, renders the geodesic equations as a set of first-order differential equations \cite{Carter}.
%
Later, the introduction of the Mino time \cite{Mino_2003} further fully decouples the equations of motions with the solutions given in terms of the elliptical functions \cite{Schmidt_2002,Fujita_2009}.
In our previous paper \cite{Wang_2022}, we have studied the null and time-like geodesics of the light and the neutral particles respectively in the exterior of Kerr-Newman  black holes.
We then obtain the solutions of the trajectories written in terms of  the elliptical integrals and the Jacobi elliptic functions \cite{Abramowitz}, in which the orbits are manifestly real functions of the Mino time and also the initial conditions can be explicitly specified \cite{Gralla_2020a}.
%
In this work, we will mainly focus on the infalling particles into the Kerr-Newman black holes in general nonequatorial motion.
Theoretical considerations, together with recent observations of structures near Sgr A* by the GRAVITY experiment \cite{Abu_2018}, indicate possible presence of a small electric charge of central supermassive black hole \cite{Zaj_2018,Zaj_2019}.
Thus, it is  of great interest to explore the geodesic dynamics in the Kerr-Newman black hole \cite{Hackmann_2013}.

Layout of the paper is as follows.
In Sec. \ref{secII}, a mini review of the time-like geodesic equations is provided with three conserved quantities of a particle, the energy, azimuthal angular momentum, and Carter constant.
The equations of motion can be recast into integral forms involving two effective potentials.
In particular,  the roots of the radial potential determine the different types of the infalling trajectories of particles to  black holes \cite{Compere_2022}.
Sec. \ref{secIII} focuses  on the parameter regime of the conserved quantities to have triple roots giving the radius of  the innermost stable spherical orbits (ISSO) \cite{Stein}.
The analytical solutions of the infalling orbits are derived for the case that the particle starts from the coordinate $r$ slightly less than that of the ISSO.
The two additional infalling orbits of interest are determined by the roots of the radial potential, consisting of a pair of complex roots and two real roots.
In Sec. \ref{secIV}, we consider bound motion with one of the real roots inside the horizon and the other outside the horizon.
In this case, the particle motion is bound by the turning point from the real root outside the horizon.
In Sec. \ref{secV} we will consider unbound motion, in which both real roots are inside the event horizon.
The exact solutions for plunging trajectories are given and illustrative examples for such trajectories are plotted.
In \ref{secVI} the conclusions are drawn.
For the clarity of notation and the completeness of the paper,  Appendixes \ref{appA} and \ref{appB} provide some of relevant formulas derived in the earlier publications \cite{Wang_2022,Li_2023}.

\section{Equation of motion for time-like geodesics}\label{secII}

We start from a summary of the equations of motion for the particle in the Kerr-Newman black hole exterior.
We work with the Boyer-Lindquist coordinates $(t,r,\theta,\phi)$. The spacetime of
the exterior of the Kerr-Newman black hole with the gravitational mass $M$, angular momentum $J$, and angular momentum per unit mass $ a=J/M$ is described by the metric
\be
ds^2=-\frac{\Delta}{\Sigma}\left(dt-a\sin^2\theta d\phi \right)^2+\frac{\sin^2\theta}{\Sigma}\left[(r^2+a^2)d\phi-a dt \right]^2+\frac{\Sigma}{\Delta}dr^2+\Sigma d\theta^2\;,
\ee
where
$\Sigma=r^2+a^2\cos^2\theta$ and $\Delta=r^2-2Mr+a^2+Q^2$.
The roots of $\Delta (r)=0$ determine outer/inner event horizons $r_{+}/r_{-}$ as
%
\be
r_{\pm}=M\pm\sqrt{M^2-(Q^2+a^2)}\;.
\ee
We assume that $0 < a^2 +Q^2 \le M^2$ throughout the paper.

For the asymptotically flat, stationary, and axial-symmetric black holes, the metric is independent of $t$ and $\phi$.
Thus, the  conserved quantities are energy $E_m$ and azimuthal angular momentum $L_m$ of the particle  along a geodesic.
These can be constructed through the four momentum  $p^\mu= m u^\mu= m \, dx^\mu /d\sigma_m $, defined in terms of the proper time $\sigma_m$ and the mass of the particle $m$ as
 \begin{align}
 E_m & \equiv -p_t ,\label{E_m}\\
L_m & \equiv p_\phi \label{L_m}\,.
 \end{align}
Additionally, another conserved quantity is the Carter constant explicitly  obtained by
 \begin{equation}
 {C}_m= \Sigma^2\left(u^{\theta}\right)^2-a^2\cos^2\theta \left({E_m}\right)^2+L_m^2\cot^2\theta+m^2a^2\cos^{2}\theta\, .\label{mathbb_C}
 \end{equation}
Together with the  time-like geodesics, $ u^\mu u_\mu=m^2$, one obtains the equations of motion
\be
\frac{\Sigma}{m}\frac{d{r}}{d\sigma_m}=\pm_r\sqrt{R_m({r})} \,,\label{r_eq_particle}
\ee
\be
\frac{\Sigma}{m}\frac{d\theta}{d\sigma_m}=\pm_{\theta}\sqrt{\Theta_m(\theta)} \,, \label{theta_eq_particle}
\ee
\be
\frac{\Sigma}{m}\frac{d\phi}{d\sigma_m}=\frac{{a}}{{\Delta}}\left[\left({r}^2+{a}^2\right)\gamma_m-{a}\lambda_m\right]-\frac{1}{\sin^{2}\theta}\left({a}\gamma_m \sin^2\theta-\lambda_m\right) \,,\label{phi_eq_particle}
\ee
\be
\frac{\Sigma}{m}\frac{d{t}}{d\sigma_m}=\frac{{r}^2+{a}^2}{{\Delta}}\left[\left({r}^2+{a}^2\right)\gamma_m-{a}\lambda_m\right]-{a}\left({a}\gamma_m \sin^2\theta-\lambda_m\right) \,,\label{t_eq_particle}
\ee
where we have normalized $E_m$, $L_m$, and $C_m$ by the mass of the particle $m$
\begin{align}
&\gamma_m\equiv\frac{E_m}{m},\hspace*{2mm}\lambda_m\equiv\frac{L_m}{m},\hspace*{2mm}\eta_m\equiv\frac{ {C}_m}{m^2}.
\end{align}
The symbols $\pm_r={\rm sign}\left(u^{r}\right)$ and $\pm_{\theta}={\rm sign}\left(u^{\theta}\right)$ are defined by the 4-velocity of the particle.
Moreover, the radial  potential $R_m({r})$ in (\ref{r_eq_particle}) and and angular potential $\Theta_m(\theta)$ in (\ref{theta_eq_particle})  are obtained as
\be
R_m({r})=\left[\left({r}^2+{a}^2\right)\gamma_m-{a}\lambda_m\right]^2-{\Delta}\left[ \eta_m+\left({a}\gamma_m-\lambda_m\right)^2+{r}^2\right]\, ,\label{Rpotential}
\ee
\be
\Theta_m(\theta)=\eta_m+{a}^2\gamma_m^2\cos^2\theta-\lambda_m^2\cot^2\theta-{a}^2\cos^2\theta \, .\label{Thetapotential}
\ee
As well known \cite{Mino_2003}, the set of equations of motion (\ref{r_eq_particle})-(\ref{t_eq_particle}) can be fully decoupled by introducing the so-called Mino time $\tau_m$ defined as
\be
\frac{dx^{\mu}}{d\tau_m}\equiv\frac{\Sigma}{m}\frac{dx^{\mu}}{d\sigma_m}\,\label{tau'}\, .
\ee

For the source point $x_{i}^{\mu}$ and observer point $x^{\mu}$, the integral forms of the equations above can be rewritten as  \cite{Gralla_2020a}
\be
\tau_m-\tau_{mi}=I_{mr}=G_{m\theta} \, ,\label{r_theta}
\ee
\be
\phi_m-\phi_{mi}= I_{m\phi}+{\lambda_{m}} G_{m\phi}\, , \label{phi}
\ee
\be
t_m-t_{mi}= I_{mt}+a^2{\gamma_{m}}G_{mt} \, ,\label{t}
\ee
where the integrals $I_{mr}$, $I_{m\phi}$, and $I_{mt}$ involve the radial potential $R_m(r)$
\be
I_{m r}\equiv\int_{r_{i}}^{r}\frac{1}{\pm_r\sqrt{R_m(r)}}dr,
\, ,\label{Imr}
\ee
\be
I_{m \phi}\equiv\int_{r_{i}}^{r}\frac{{a\left[\left(2Mr-Q^2\right)\gamma_{m}-a\lambda_{m}\right]}}{\pm_r\Delta\sqrt{R_m(r)}}dr,\label{Imphi}
\ee
\be
I_{mt}\equiv\int_{r_{i}}^{r}\frac{{r^2\gamma_{m}\Delta+(2Mr-Q^2)\left[\left(r^2+a^2\right)\gamma_{m}-a{\lambda_{m}}\right]}}{\pm_r\Delta\sqrt{R_m(r)}}dr\, .\label{Imt}
\ee
The angular integrals are
\be
G_{m \theta}\equiv\int_{\theta_{i}}^{\theta}\frac{1}{\pm_{\theta}\sqrt{\Theta_m(\theta)}}d\theta \, ,\label{Gmtheta}
\ee
\be
G_{m \phi}\equiv\int_{\theta_{i}}^{\theta}\frac{\csc^2\theta}{\pm_{\theta}\sqrt{\Theta_m(\theta)}}d\theta \, ,\label{Gmphi}
\ee
\be
G_{mt}\equiv\int_{\theta_{i}}^{\theta}\frac{\cos^2\theta}{\pm_{\theta}\sqrt{\Theta_m(\theta)}}d\theta \, . \label{Gmt}
\ee
%

{
The radial potential $R_m(r)$ is given by a quartic polynomial and the types of its roots play essential roles in the classification of different orbits.
In the previous work \cite{Wang_2022,Li_2023}, we have shown the exact solutions to some of the cases of both null and time-like geodesics.
For the present work, we will mainly focus on the infalling orbits of the bound and the unbound motion at the black hole exterior.
In this context, we will consider three types of such orbits in the subsequent sections.
%
The discussion of the angular potential $\Theta(\theta)$ and the integrals involved has been presented in Ref. \cite{Wang_2022}. For the sake of completeness, we will provide a short summary  in  Appendix \ref{appA}.
}

{
Before ending this section, let us introduce a few notations that will be used in the subsequent sections.
Related to $R_m(r)$ we define the integrals
\be
I_{n}\equiv\int_{r_{i}}^{r} r^n\sqrt{\frac{1-\gamma_m^2}{R_m(r)}}\,dr
\equiv i I^U_{n}\;,\;n=1,2\label{I_n} \,,
\ee
\be
I_{\pm}\equiv\int_{r_{i}}^{r}\frac{1}{\left(r-r_{\pm}\right)}
\sqrt{\frac{1-\gamma_m^2}{R_m(r)}}\,dr\equiv i I^U_{\pm}\,.\label{Ipm}
\ee
In terms of $I_1$, $I_2$, and $I_\pm$ we can rewrite (\ref{Imphi}) and (\ref{Imt}) as follows
\be
I_{m\phi}(\tau_m)=\frac{\gamma_m}{\sqrt{1-\gamma_m^2}}\frac{2Ma}{r_{+}-r_{-}}\left[\left(r_{+}-\frac{a\left(\frac{\lambda_m}{\gamma_m}\right)+Q^2}{2M}\right)I_{+}(\tau_m)-\left(r_{-}-\frac{a\left(\frac{\lambda_m}{\gamma_m}\right)+Q^2}{2M}\right)I_{-}(\tau_m)\right]\;,
\label{Imphi_0}
\ee
\begin{align}
&I_{mt}(\tau_m)=\frac{\gamma_m}{\sqrt{1-\gamma_m^2}}\left\lbrace\frac{4M^2}{r_{+}-r_{-}}\left[\left(r_{+}-\frac{Q^2}{2M}\right)\left(r_{+}-\frac{a\left(\frac{\lambda_m}{\gamma_m}\right)+Q^2}{2M}\right)I_{+}(\tau_m)\right.\right.\notag\\
&\quad\quad \quad\quad\quad\quad\quad\left.\left.-\left(r_{-}-\frac{Q^2}{2M}\right)\left(r_{-}-\frac{a\left(\frac{\lambda_m}{\gamma_m}\right)+Q^2}{2M}\right)I_{-}(\tau_m)\right]+2MI_{1}(\tau_m)+I_{2}(\tau_m)
\right\rbrace\notag\\
&\quad\quad \quad\quad\quad\quad\quad+\left(4M^2-Q^2\right)\gamma_{m}\tau_m\;.
\label{Imt_0}
\end{align}
}

\section{INSPIRAL ORBITS IN BOUND MOTION} \label{secIII}

\begin{figure}[h]
 \centering
 \includegraphics[scale=0.5]{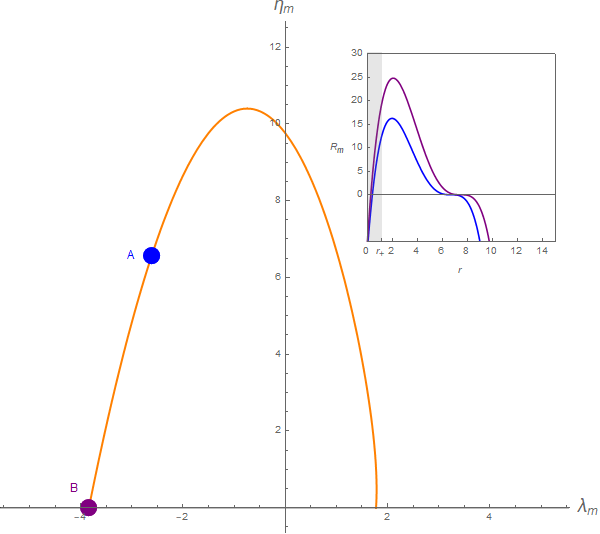}
 \caption{
{The main graphics shows the parametric plot of $\lambda_m(r_{\rm isso})$ versus $\eta_m(r_{\rm isso})$. The triple roots $r_{\rm isso}$ are the solutions of the equations $R''_{m}(r)=R'_{m}=R_m (r)=0$.
The inset illustrates the behavior of the radial potential $R_m$ with the parameters located at A and B. The case of B has $\eta_m=0$, which is an example of equatorial motion. }
\label{Rm_r_triple}}
\end{figure}

According to previous studies of radial potential \cite{Wang_2022}, by examining different ranges of the parameters $\lambda_m$ and $\eta_m$ for bound motion ($\gamma_m <1$), it is clear that there are two distinct categories of infalling motion that traverse the horizon and enter the black hole.
One is that the particle starts from $r_i \le r_{\rm isso}$, where the radius of the ISSO orbit $r_{\rm isso}$ is within the parameters located at {A and B} in Fig. \ref{Rm_r_triple},
spirals and then plunge into the horizon of the black holes \cite{Ori_2000,O_Shaughnessy_2003}.
The other one is starting from $r_i < r_{m4}$, with the parameters of {C and D} in Fig. \ref{Bound_Plunge}, and plunge through the horizon of the black holes.
%

%

We first consider the particle starts from $r_i \le r_{\rm isso}$.
The solutions along the $r$ direction can be obtained from the inversion of (\ref{r_theta}) with the integral $I_{mr}$ in (\ref{Imr}), where the radial potential (\ref{Rpotential}) is given in the case of the triple root located at the ISSO radius, namely $r_{m2}=r_{m3}=r_{m4}=r_{\rm isso}$, and the initial $r$ is set at $r_i\le r_{\rm isso}$.
So, from the integral (\ref{r_theta}) and (\ref{Imr}),
we can have the Mino time $\tau_{m}$ as the function of $r$
 \begin{equation}\label{tau_r 1}
 \tau^I_{m}(r)=\frac{- 2}{(r_{\rm isso}-r_{m1})\sqrt{1-\gamma_{m}^2}}\Bigg[ \sqrt{\frac{r-r_{m1}}{r_{\rm isso}-r}}-\sqrt{\frac{r_i-r_{m1}}{r_{\rm isso}-r_i}}\Bigg]\, .
 \end{equation}
The particle moves toward the horizon with $\nu_{r_i}=-1$. Thus, the inverse of (\ref{tau_r 1}) leads to
%
%
\be
r^I(\tau_m)=\frac{r_{m1}+r_{\rm isso}\left[X^{I}(\tau_m)]^2\right.}{1+{\left[X^{I}(\tau_m)]^2\right.}}\;,\label{r_tau}
\ee
where
\be
X^{I}(\tau_m)=\frac{\sqrt{1-\gamma_m^2}(r_{\rm isso}-r_{m1})}{2}\tau_m-\sqrt{\frac{r_i-r_{m1}}{r_{\rm isso}-r_i}} \label{XI} \, .
\ee
The solution (\ref{r_tau}) of the coordinate $r$ involves the triple root $r_{\rm isso}$ of the radial potential, which can be determined as follows.

From the double root solutions $R(r)=R'(r)=0$ \cite{Wang_2022}, two constants of motion in the case of spherical orbits are given by
%
%
\be
\lambda_{\rm mss}=\frac{\left[r_{\rm mss}\left(Mr_{\rm mss}-Q^2\right)-a^2M\right]\gamma_m-\Delta\left(r_{\rm mss}\right)\sqrt{r_{\rm mss}^2\left(\gamma_m^2-1\right)+Mr_{\rm mss}}}{a\left(r_{\rm mss}-M\right)}\;, \label{tilde_lambda_m}
\ee
\begin{align}
&{\eta}_{\rm mss}=\frac{r_{\rm mss}}{a^2\left(r_{\rm mss}-M\right)^2}
\Big\{ r_{\rm mss}\left(M r_{\rm mss}-Q^2\right)\left(a^2+Q^2-Mr_{\rm mss}\right)\gamma_m^2\Big.
\notag\\
&\quad\quad\quad\quad\quad+2\left(Mr_{\rm mss}-Q^2\right)\Delta\left(r_{\rm mss}\right)\gamma_m\sqrt{r_{\rm mss}^2\left(\gamma_m^2-1\right)+Mr_{\rm mss}}\notag\\
& \Big.
\quad\quad\quad\quad\quad\left.+\left[a^2\left(Mr_{\rm mss}-Q^2\right)-\left(\Delta\left(r_{\rm mss}\right)-a^2\right)^2\right]\left[r_{\rm mss}\left(\gamma_m^2-1\right)+M\right]\Big\} \right.\;.  \label{tilde_eta_m}
\end{align}
The subscript "ss" means the spherical orbits with $s=\pm$, which denotes the two types of motion with respect to the relative sign between the black hole's spin and the azimuthal angular of the particle (see Section III C of \cite{Wang_2022}).
Together with the two relations above, an additional equation from $R''(r)=0$ determines the triple roots.
We have found the radius of $r_{\rm isso}$ satisfying the following equation
%

{
\begin{equation}\label{r_isso_noneq}
    -M r_{\rm isso}^5\Delta\left(r_{\rm isso}\right)+4\left(M r_{\rm isso}^3-Q^2r_{\rm isso}^2+a^2\eta_{\rm isso}-as\sqrt{\Gamma_{\rm ms}}\right)^2=0\,,
\end{equation}
%
where
\be
\Gamma_{\rm ms}=r_{\rm isso}^4\left(M r_{\rm isso}-Q^2\right)
-\eta_{\rm isso}\left[r_{\rm isso}\left(r_{\rm isso}-3M\right)+2Q^2\right]r_{\rm isso}^2+a^2\eta_{\rm isso}^2 \,. \label{|gamma_noneq}
\ee

We proceed by evaluating the coordinates $\phi_m(\tau_m)$ and $t_m(\tau_m)$ using (\ref{phi}) and (\ref{t}), which involve not only the angular integrals  $G_{m\phi}$ and $G_{mt}$, but also the radial integrals (\ref{Imphi}) and (\ref{Imt}).
With the help of (\ref{Imphi_0}) and (\ref{Imt_0}), we first rewrite (\ref{Imphi}) and (\ref{Imt}) as
\be
I^{I}_{m\phi}(\tau_m)=\frac{\gamma_m}{\sqrt{1-\gamma_m^2}}\frac{2Ma}{r_{+}-r_{-}}\left[\left(r_{+}-\frac{a\left(\frac{\lambda_m}{\gamma_m}\right)+Q^2}{2M}\right)I_{+}^{I}(\tau_m)-\left(r_{-}-\frac{a\left(\frac{\lambda_m}{\gamma_m}\right)+Q^2}{2M}\right)I_{-}^{I}(\tau_m)\right]\label{Imphi2}\,,\\
\ee
\begin{align}
&I^{I}_{mt}(\tau_m)=\frac{\gamma_m}{\sqrt{1-\gamma_m^2}}\left\lbrace\frac{4M^2}{r_{+}-r_{-}}\left[\left(r_{+}-\frac{Q^2}{2M}\right)\left(r_{+}-\frac{a\left(\frac{\lambda_m}{\gamma_m}\right)+Q^2}{2M}\right)I_{+}^{I}(\tau_m)\right.\right.\notag\\
&\quad\quad \quad\quad\quad\quad\left.\left.-\left(r_{-}-\frac{Q^2}{2M}\right)\left(r_{-}-\frac{a\left(\frac{\lambda_m}{\gamma_m}\right)+Q^2}{2M}\right)I_{-}^{I}(\tau_m)\right]+2MI_{1}^{I}(\tau_m)+I_{2}^{I}(\tau_m) \right\rbrace\notag\\
&\quad\quad \quad\quad\quad\quad+\left(4M^2-Q^2\right)\gamma_{m}\tau_m\;. \label{Imt2}
\end{align}
For the present case of the triple roots, the calculation of the integrals is straightforward and one can express  $I_{n}^{I}$ and  $I_{\pm}^{I}$ in terms of elementary functions,
%
\be
I_{\pm}^{I}(\tau_m)=\frac{\sqrt{1-\gamma_m^2}}{r_{\rm isso}-r_{\pm}}\tau_m+\frac{1}{\sqrt{(r_{\pm}-r_{m1})\left(r_{\rm isso}-r_{\pm}\right)^3}}
\tanh^{-1}
\sqrt{\frac{(r_{\pm}-r_{m1})(r_{\rm isso}-r^{I}(\tau_{m}))}{(r_{\rm isso}-r_{\pm})(r^{I}
(\tau_{m})-r_{m1})}}
-\mathcal{I}_{\pm_i}^{I}\,,\label{Ipmisso}
\ee
\be
I_{1}^{I}(\tau_m)=\sqrt{1-\gamma_m^2}r_{\rm isso}\tau_m+2\tan^{-1}\sqrt{\frac{r_{\rm isso}-r^{I}(\tau_m)}{r^{I}(\tau_m)-r_{m1}}}-\mathcal{I}_{1_i}^{I}\,,
\ee
\be
I_{2}^{I}(\tau_m)=\frac{r_{I}(\tau_m)(r_{m1}-r_{\rm isso})+r_{\rm isso}(3r_{\rm isso}-r_{m1})}{2}\tau_m+(r_{m1}+3 r_{\rm isso})\tan^{-1}\sqrt{\frac{r_{\rm isso}-r^{I}(\tau_m)}{r^{I}(\tau_m)-r_{m1}}}-\mathcal{I}_{2_i}^{I}\,.
\ee
%
It is worthwhile to mention that
$\mathcal{I}_{\pm_i}^{I}$,  $\mathcal{I}_{1_i}^{I}$, $\mathcal{I}_{2_i}^{I}$ are obtained  by evaluating ${I}_{\pm}^{I}$,  ${I}_{1}^{I}$, ${I}_{2}^{I}$ at $r=r_i$ of the initial condition, that is,
${{I}_{\pm}^{I}(0)={I}_{1}^{I}(0)={I}_{2}^{I}(0)=0}$.
The solutions of $\phi^I(\tau_m)$ and $t^I(\tau_m)$ can be constructed from $I_{m\phi}$ (\ref{Imphi}), $G_{m\phi}$ (\ref{Gmphi}) and $I_{mt}$ (\ref{Imt}) and $G_{mt}$ (\ref{Gmt}) through (\ref{phi}) and (\ref{t}).
Together with the solutions along  $r$ and  $\theta$ directions in (\ref{r_tau}) and (\ref{theta_tau}), they are infalling motions of the general nonequatorial orbits in the Kerr-Newman exterior.
An illustrative example is shown in Fig. \ref{IO_1_nonequa} with the parameters of the case A in  Fig. \ref{Rm_r_triple}.
The particle starts by inspiraling around $r_{\rm isso}$ and then plunges into the black hole's horizon.
%
%

\begin{figure}[h]
 \centering
 \includegraphics[scale=0.5]{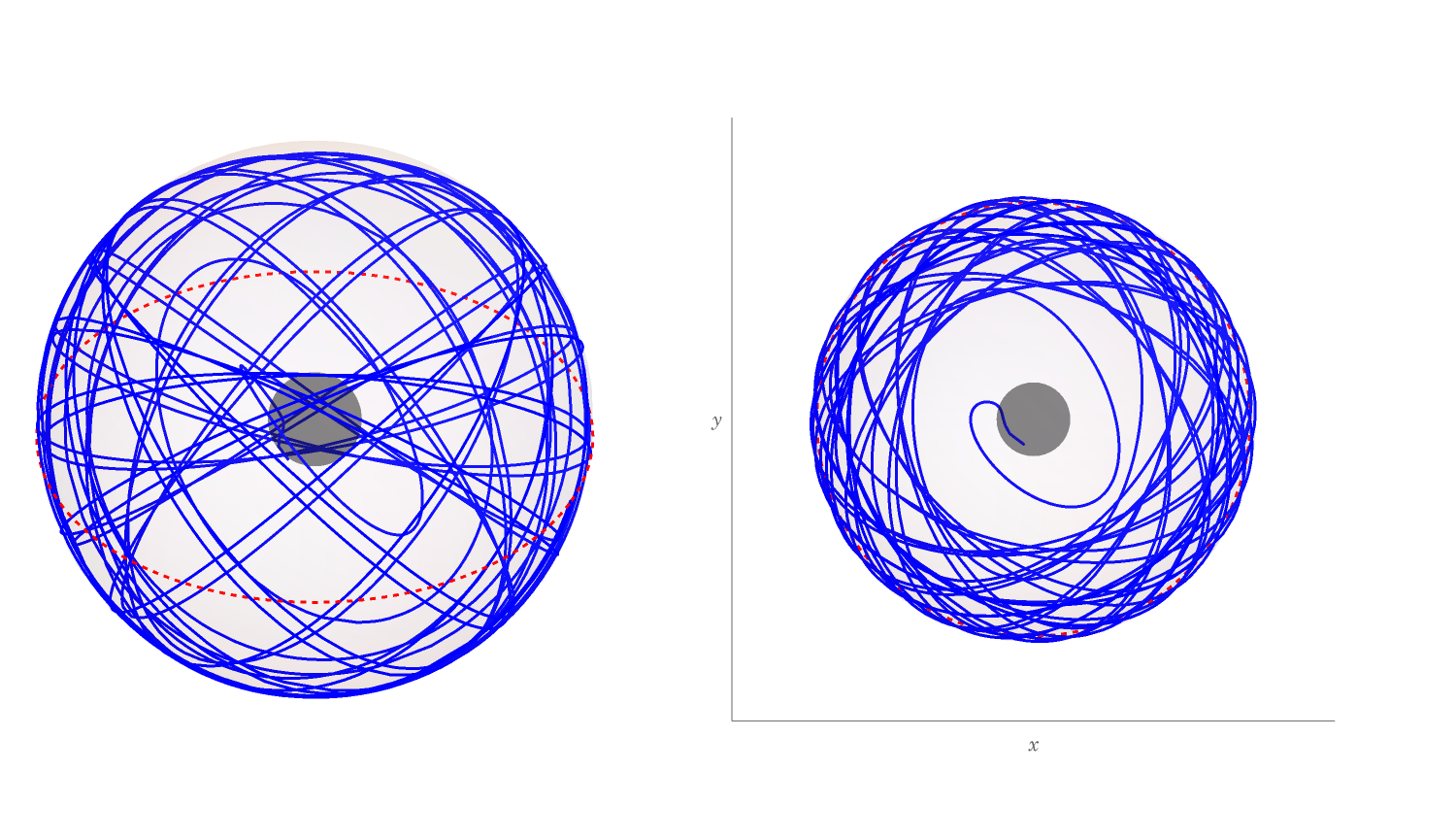}
 \caption{
{An illustrative example of nonequatorial orbit with parameters A of Fig. \ref{Rm_r_triple}.
In this case, the particle starts from $r_i \le r_{\rm isso}$ and inspirals into the black hole after many azimuthal and longitudinal revolutions.
From the top view one notices the very different time scales of the spiralling and plunging phases }
\label{IO_1_nonequa}}
 \end{figure}

For the particle initially at $r_i=r_{\rm isso}$, the solution (\ref{r_tau}) gives $r (\tau)=r_{\rm isso}$ obtained from $X^I \rightarrow - \infty$ in (\ref{XI}) when $r_i \rightarrow r_{\rm isso}$, meaning that the particle will stay in spherical motion with the $r_{\rm isso}$ radius.  However, for $r_i < r_{\rm isso}$ of our interest, as $r$ reaches the outer horizon $r_+$,
it takes finite Mino time $\tau_m$.
Nevertheless, because of the $\tanh^{-1}$ function in (\ref{Ipmisso}), $I_{\pm}^{\rm isso} \rightarrow \infty$ as $r  \rightarrow r_+$, giving the coordinate time $t \rightarrow \infty$ and the azimuthal angle $\phi \rightarrow \infty$ observed in the asymptotical flat regime. 
{The above expressions can be further reduced to the Kerr black hole case by sending $Q\rightarrow0$ in \cite{DYS}.
As for the contributions to the evolution of the angle $\phi$ in (\ref{phi}) from the integrals involving the radial potential $R_m(r)$, one can write the $\tanh^{-1}$ function  in $I_{\pm}^{I}(\tau_m)$  by $\log$ function through $\tanh^{-1}(x)=\frac{1}{2} \frac{\log (1+x)}{\log (1-x)}$. Then, the remaining terms in (\ref{Imphi2}) directly proportional to $\tau_m$ are the same in \cite{DYS}. Together with (\ref{GmphiA}) of the integrals involving the $\theta$ potential $\Theta_m(\theta)$,  the variables $z_1$ and $z_2$ defined in \cite{DYS} can be related to the roots of $\Theta_m(\theta)$ by
$z_1^2=u_{m+}$ and $z_2^2=u_{m-} a^2 (1-\gamma_m^2)$, leading to $k_z^2=\frac{u_{m+}}{u_{m-}}$, which gives the same expression in  \cite{DYS} from (\ref{GmphiA}).}

One of the interesting  cases is considering the equatorial motion by taking $\theta=\frac{\pi}{2}$ and $\eta_{m}\rightarrow0 $ limits in the results above.
The particle starts from the coordinate $r$ slightly less than the radius of innermost circular motion $r_{\rm isco}$.
In particular, $G_{m\phi}=\tau_m$ and the solution of $\phi^I_m$ in  (\ref{phi}) simplifies to \cite{Wang_2022,Li_2023}
\be
\phi^I_m\left(r\right)=I^{I}_{m\phi}\left(\tau_{m}\left(r\right)\right)+\lambda_{m}\tau^I_{m}\left(r\right)+\phi^I_{mi}\,,
\ee
where $I^{I}_{m\phi}$ is given by (\ref{Imphi2}).
In addition, one can replace Mino time $\tau_m$ by the coordinate $r$ through Eq. (\ref{tau_r 1}).
Then the infalling solution of the angle $\phi_m$
on the equatorial plane
can be expressed as a function of $r$,
\begin{align}
&\phi^I_m(r)=-2\sqrt{\frac{r-r_{m1}}{\left(1-\gamma_{m}^2\right)\left(r_{\rm isco}-r\right)}}\frac{r_{\rm isco}^2\lambda_{m}+\left(2M r_{\rm isco}-Q^2\right)\left(a\gamma_{m}-\lambda_{m}\right)}{\left(r_{\rm isco}-r_{+}\right)\left(r_{\rm isco}-r_{-}\right)\left(r_{\rm isco}-r_{m1}\right)}\notag\\
&-\frac{2}{r_{+}-r_{-}}\frac{\left(2M a\gamma_{m}-r_{-}\lambda_{m}\right)r_{+}-Q^2\left(a\gamma_{m}-\lambda_{m}\right)}{\left(r_{\rm isco}-r_{+}\right)\sqrt{\left(1-\gamma_{m}^2\right)\left(r_{+}-r_{m1}\right)\left(r_{\rm isco}-r_{+}\right)}}\tanh^{-1}\sqrt{\frac{\left(r_{+}-r_{m1}\right)\left(r_{\rm isco}-r\right)}{\left(r_{\rm isco}-r_{+}\right)\left(r-r_{m1}\right)}}\notag\\
&+\frac{2}{r_{+}-r_{-}}\frac{\left(2M a\gamma_{m}-r_{+}\lambda_{m}\right)r_{-}-Q^2\left(a\gamma_{m}-\lambda_{m}\right)}{\left(r_{\rm isco}-r_{-}\right)\sqrt{\left(1-\gamma_{m}^2\right)\left(r_{-}-r_{m1}\right)\left(r_{\rm isco}-r_{-}\right)}}\tanh^{-1}\sqrt{\frac{\left(r_{-}-r_{m1}\right)\left(r_{\rm isco}-r\right)}{\left(r_{\rm isco}-r_{-}\right)\left(r-r_{m1}\right)}}\, .\label{phi_equ}
\end{align}
Likewise, for the equatorial orbits, Eq. (\ref{t}) with $G_{mt}=0$ gives
\be
t^I_m\left(r\right)=I^{\rm I}_{mt}\left(\tau_{m}\right)+t^I_{mi}\,,
\ee
where $I^{I}_{mt}$ has been calculated in (\ref{Imt2}). Substituting $\tau^I_m$ to replace $r$,  we find
\begin{align}
&t^I_m\left(r\right)=-\gamma_{m}\sqrt{\frac{\left(r-r_{m1}\right)\left(r_{\rm isco}-r\right)}{1-\gamma_{m}^2}}+\frac{\gamma_{m}\left(r_{m1}+3r_{\rm isco}+4M\right)}{\sqrt{1-\gamma_{m}^2}}\tan^{-1}\sqrt{\frac{r_{\rm isco}-r}{r-r_{m1}}}\notag\\
&-2\sqrt{\frac{r-r_{m1}}{\left(1-\gamma_{m}^2\right)\left(r_{\rm isco}-r\right)}}\frac{r_{\rm isco}^2\left(r_{\rm isco}^2+a^2\right)\gamma_{m}+\left(2M r_{\rm isco}-Q^2\right)a\left(a\gamma_{m}-\lambda_{m}\right)}{\left(r_{\rm isco}-r_{+}\right)\left(r_{\rm isco}-r_{-}\right)\left(r_{\rm isco}-r_{m1}\right)}\notag\\
&-\frac{2\left(2M r_{+}-Q^2\right)}{r_{+}-r_{-}}\frac{2M\gamma_{m}r_{+}-\left(a\lambda_{m}+Q^2\gamma_{m}\right)}{\left(r_{\rm isco}-r_{+}\right)\sqrt{\left(1-\gamma_{m}^2\right)\left(r_{+}-r_{m1}\right)\left(r_{\rm isco}-r_{+}\right)}}\tanh^{-1}\sqrt{\frac{\left(r_{+}-r_{m1}\right)\left(r_{\rm isco}-r\right)}{\left(r_{\rm isco}-r_{+}\right)\left(r-r_{m1}\right)}}    \notag\\
&+\frac{2\left(2M r_{-}-Q^2\right)}{r_{+}-r_{-}}\frac{2M\gamma_{m}r_{-}-\left(a\lambda_{m}+Q^2\gamma_{m}\right)}{\left(r_{\rm isco}-r_{-}\right)\sqrt{\left(1-\gamma_{m}^2\right)\left(r_{-}-r_{m1}\right)\left(r_{\rm isco}-r_{-}\right)}}\tanh^{-1}\sqrt{\frac{\left(r_{-}-r_{m1}\right)\left(r_{\rm isco}-r\right)}{\left(r_{\rm isco}-r_{-}\right)\left(r-r_{m1}\right)}}\,.
\label{t_equ}
\end{align}
As for the initial conditions one can determine $\phi^I_{mi}$ and $t^I_{mi}$ by  $I^I_{m\phi}\left(\tau^I_{m}\left(r\right)\right)+\lambda_{m}\tau^i_{m}\left(r\right)$ and $I^I_{mt}\left(\tau_{m}\right)$ vanishing at the initial $r_i$.
The corresponding trajectories are shown in Fig. \ref{IO_1_equa}, with the additional parameter $Q$ apart from $a$ of the black holes.
This generalizes the solution in \cite{{mummery2022inspirals}} for the Kerr black holes, where the particle starts from $r \lesssim r_{\rm isco}$ at $t_{m}(r)=-\infty$  and  inspirals to the event horizon.
{In the limit of $Q\rightarrow 0$  where $r_{m1}\rightarrow 0$, $1-\gamma_m^2 \rightarrow {2M}/{3 r_{\rm isco}}$, and $(r_{\rm isco}-r_+)(r_{\rm isco}-r_-) \rightarrow r^2_{\rm isco}-2 M r_{\rm isco}+a^2$, the first term of Eq.(\ref{phi_equ}) reduces to the corresponding term  $\sqrt{\frac{r}{r_{\rm isco}-r}}$ of $\phi^I_m(r)$ in \cite{mummery2022inspirals}. Together with $t_{\pm}=\sqrt{\frac{r_{\pm}}{r_{\rm isco}-r_{\pm}}}$ defined in \cite{mummery2022inspirals}, one can make the replacement
\begin{equation}
\frac{1}{r_+-r_-} \rightarrow \frac{1}{t^2_+-t^2_-} \frac{r_{\rm isco}}{r^2_{\rm isco}-2 M r_{\rm isco}+a^2} \, ,
\end{equation}}
{to reproduce the $\tanh^{-1}$  terms.}

One of the  limiting cases that can significantly simplify the above expressions is to consider the extremal limit of the Kerr black hole, $a\rightarrow M$. For $Q\rightarrow0$ giving $r_{m1}=0$,  and for the extremal black holes, the ISCO radius for direct orbits is on the event horizon.
Here we focus on the extremal retrograde motion with $r_{\rm isco}=9M$, $\lambda_{m}=-22\sqrt{3}M/9$ and $\gamma_{m}=5\sqrt{3}/9$.
%
{Notice that  in the extremal black holes, $r_{+}$ and $r_{-}$ collapse into
the same value.}
Then the solutions can be reduced into the known ones \cite{mummery2022inspirals,Mummery_2023}
\be \label{phi_r_KN_e}
\phi^I_m\left(r\right)=-\frac{2\sqrt{2}}{3}\frac{r^\frac{3}{2}}{(r-M)\sqrt{9M-r}}\,,
\ee
\begin{align}
&t^I_m\left(r\right)=\sqrt{\frac{(9M-r)r}{2}}\left(\frac{4M-5r}{r-M}\right)-\frac{117\sqrt{2}}{2}M\sqrt{\frac{r}{9M-r}}\notag\\
&\quad\quad\quad\quad+\frac{155\sqrt{2}}{2}M\tan^{-1}\sqrt{\frac{9M-r}{r}}-4M\tanh^{-1}\sqrt{\frac{9M-r}{8r}}\, .
\end{align}

\begin{figure}[h]
 \centering
 \includegraphics[scale=0.4]{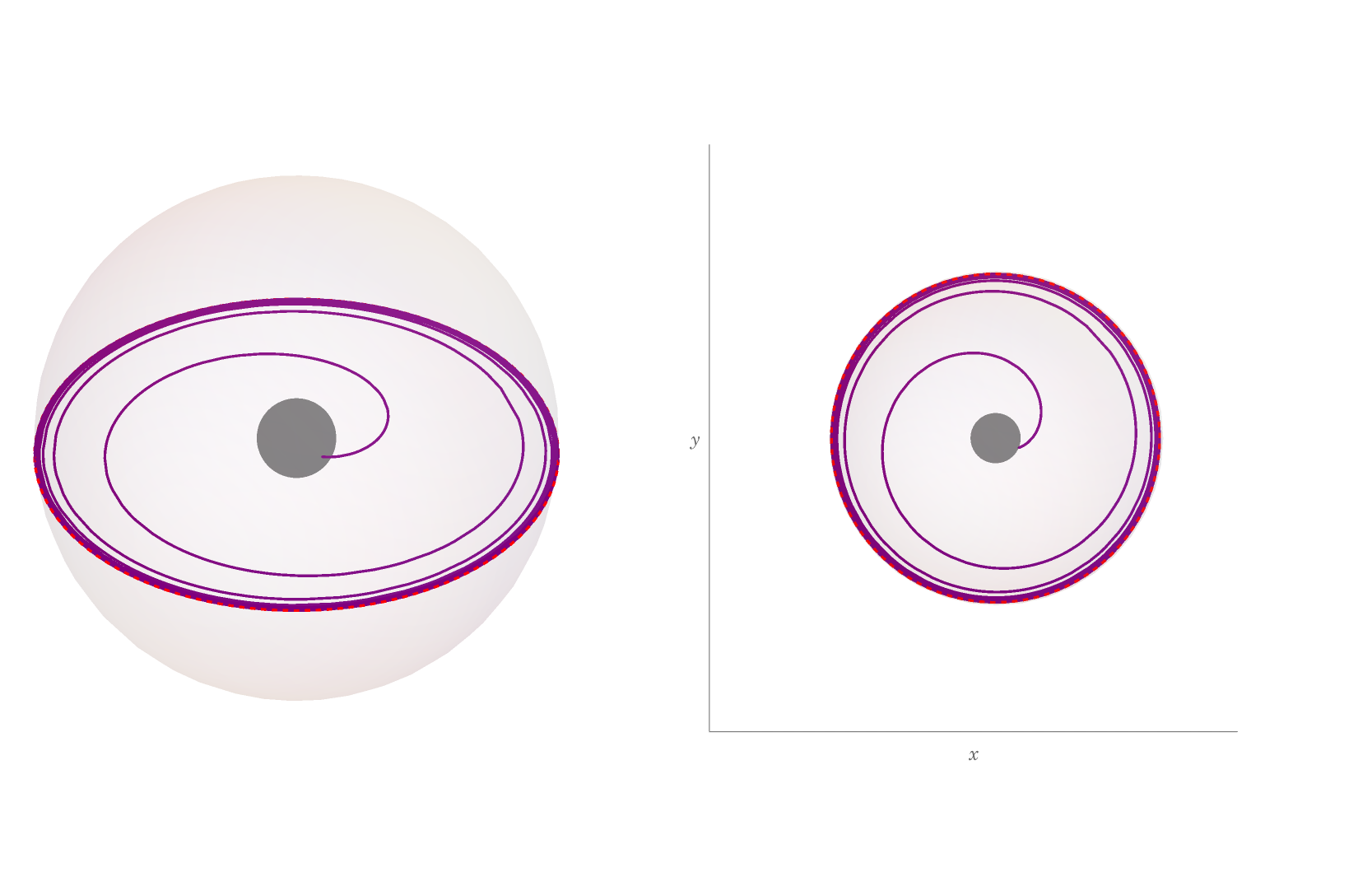}
 \caption{
{Illustration of the orbit on the equatorial plane with the parameters of B in Fig. \ref{Rm_r_triple}. The particle starts from $r_i < r_{\rm isso}$ and inspirals into the black hole horizon.}
  \label{IO_1_equa}}
 \end{figure}

Another limiting case is considering the Reissner Nordstr$\ddot{o}$m (RN) black hole.
Since $a\rightarrow0$ leading to the spherically symmetric metric,
 the general motion can be studied by considering equatorial motions. The coefficients of the $\tanh^{-1}$ terms of the above expressions (\ref{phi_equ})  all vanish.
 The expressions of $\phi^I_m\left(r\right)$ and $t^I_m\left(r\right)$  can be simplified as
\be
\phi^I_{m}\left(r\right)=-\frac{2\lambda_{m}}{r_{\rm isco}-r_{m1}}\sqrt{\frac{r-r_{m1}}{(1-\gamma_{m}^2)(r_{\rm isco}-r)}}\, , \label{phi m in RN}
\ee
\begin{align}
t^I_{m}\left(r\right)=&-\gamma_{m}\sqrt{\frac{\left(r-r_{m1}\right)\left(r_{\rm isco}-r\right)}{1-\gamma_{m}^2}}+\frac{\gamma_{m}\left(r_{m1}+3r_{\rm isco}+4M\right)}{\sqrt{1-\gamma_{m}^2}}\tan^{-1}\sqrt{\frac{r_{\rm isco}-r}{r-r_{m1}}}\notag\\
&-2\sqrt{\frac{r-r_{m1}}{\left(1-\gamma_{m}^2\right)\left(r_{\rm isco}-r\right)}}\frac{r_{\rm isco}^4\gamma_{m}}{\left(r_{\rm isco}-r_{+}\right)\left(r_{\rm isco}-r_{-}\right)\left(r_{\rm isco}-r_{m1}\right)}\notag\\
&-\frac{2}{r_{+}-r_{-}}\frac{\left(2Mr_{+}-Q^2\right)^2\gamma_{m}}{\left(r_{\rm isco}-r_{+}\right)\sqrt{\left(1-\gamma_{m}^2\right)\left(r_{+}-r_{m1}\right)\left(r_{\rm isco}-r_{+}\right)}}\tanh^{-1}\sqrt{\frac{\left(r_{+}-r_{m1}\right)\left(r_{\rm isco}-r\right)}{\left(r_{\rm isco}-r_{+}\right)\left(r-r_{m1}\right)}}\notag\\
&+\frac{2}{r_{+}-r_{-}}\frac{\left(2Mr_{-}-Q^2\right)^2\gamma_{m}}{\left(r_{\rm isco}-r_{-}\right)\sqrt{\left(1-\gamma_{m}^2\right)\left(r_{-}-r_{m1}\right)\left(r_{\rm isco}-r_{-}\right)}}\tanh^{-1}\sqrt{\frac{\left(r_{-}-r_{m1}\right)\left(r_{\rm isco}-r\right)}{\left(r_{\rm isco}-r_{-}\right)\left(r-r_{m1}\right)}}\, .\label{t m in RN}
\end{align}
Further simplification occurs in the extremal limit. For $M=\pm Q$ in the RN black holes, $r_{\pm}=M$,  and with $r_{\rm isco}=4M$, $r_{m1}=4M/5$, $\lambda_{m}=2\sqrt{2}M$ and $\gamma_{m}=3\sqrt{6}/8$, (\ref{phi m in RN}) and (\ref{t m in RN}) can have such a simple form
\be
\phi^I_{m}\left(r\right)=-2\sqrt{\frac{5r-4M}{4M-r}}\, ,
\ee
\begin{align}
t^I_{m}\left(r\right)=&-3\sqrt{\frac{3(4M-r)(r-4M/5)}{5}}+\frac{252\sqrt{15}}{25}M\tan^{-1}\sqrt{\frac{4M-r}{r-4M/5}}\notag\\
&-32M\sqrt{\frac{5r-4M}{12M-3r}}-\frac{(2M^2-1)^2}{(M-r)M^3}\sqrt{\frac{(4M-r)(5r-4M)}{3}}\notag\\
&-\frac{4(2M^2-1)}{M^3}\tanh^{-1}\sqrt{\frac{4M-r}{15r-12M}}\, .
\end{align}
Finally, in the limits of $Q\rightarrow0$ and $a\rightarrow0$ we have the case of Schwarzschild black hole, with  the spherical symmetric metric.
The horizons become $r_{+}\rightarrow 2M$ and $r_- \rightarrow 0$, and the general motion can be considered in the equatorial plane.
Thus, with the further ISCO inputs in this case, $r_{\rm isco}=6M$, $\lambda_{m}=2\sqrt{3}M$, and $\gamma_{m}=2\sqrt{2}/3$, the solutions become as simple as
\be
\phi^I_m\left(r\right)=-2\sqrt{3}\sqrt{\frac{r}{6M-r}}\;,\label{phi case1 in Schwarzschild}
\ee
\begin{align}
t^I_m\left(r\right)&=\frac{864\sqrt{2}M}{25}\sqrt{\frac{r}{6M-r}}-2\sqrt{2}\sqrt{(6M-r)r}\notag\\
&+44\sqrt{2}M\tan^{-1}\sqrt{\frac{6M-r}{r}}-4M\tanh^{-1}\sqrt{\frac{6M-r}{2r}}\;.\label{t case1 in Schwarzschild}
\end{align}
We then recover the results of two recent publications \cite{mummery2022inspirals,Mummery_2023}.
\begin{figure}[h]
\centering
\includegraphics[width=0.8\columnwidth=1.0]{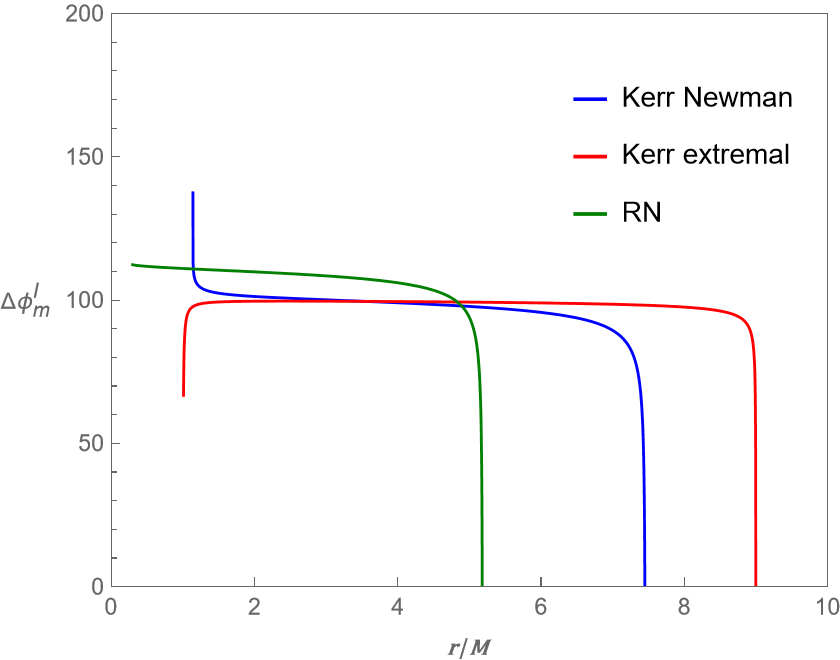}
\caption{
{
The plots show the variation of azimuthal angle  $\Delta\phi_m^{I}\equiv\phi_{m}^{I}-\phi^I_{mi}$ as a function of $r$ for the inspire bound motion for various black-hole models, including Kerr-Newman ($Q=0.7M, a=0.7M, \lambda_m=-3.84 M, \eta_m=0 M$), Kerr-extremal ($Q=0 M, a=M, \lambda_m=-\frac{22 \sqrt{3}}{9}M, \eta_m=0 M$) and RN ($Q=0.7 M, a=0 M, \lambda_m=3.2 M, \eta_m=0 M$).
In the figure, the blue curve and red curves illustrate the direct and retrograde orbits, respectively. See the text for more discussion.
}\label{phi_r_ISCO}
}
 \end{figure}

 {It is then of great interest to plot and observe how the azimuthal angle $\phi$ depends on the evaluation of coordinate $r$, as illustrated for a few exemplary cases in Fig. \ref{phi_r_ISCO}.
 In particular,  when the motion approaches the horizon, the angle $\phi$ diverges for the spinning black holes  whereas  it remains finite  for non-spinning black holes.
 From Eq.  (\ref{phi m in RN}) for RN black holes and Eq. (\ref{phi case1 in Schwarzschild}) for Schwarzschild black holes,   $\phi$ smoothly changes across the horizon.
 Thus,  another usefulness of the obtained result in (\ref{phi_equ}) is to examine the behavior of $\phi$ across the horizon.   The divergence in the azimuthal angle arises from the $\tanh^{-1}$ terms and the straightforward calculations shows that
 $\phi_m^I \sim \ln (r-r_+)$. However, for the extremal Kerr-Newman  black holes where $r_+=r_-=M$, the additional divergence in the coefficients of the $\tanh^{-1}$ terms in  (\ref{phi_equ}) shifts the  leading order divergence into that of $\phi_m^I \sim 1/ (r-M)$ in (\ref{phi_r_KN_e}), apart from  the $\ln (r-M)$ divergence.  However, for the extremal Kerr black holes,  $\phi_m^I \sim 1/ (r-M)$ from  (\ref{phi_r_KN_e}), where the  $\ln (r-M)$ divergence disappears.
%
{The dramatic difference in the behavior of the azimuthal angle $\phi$ across the horizon has been found
on equatorial infalling trajectories in Kerr black holes \cite{mummery2022inspirals,Mummery_2023}. The same types of phenomena are also seen on infalling trajectories of general nonequatorial orbits in Kerr-Newman black holes. Notice that in all cases, the coordinate time $t\rightarrow \infty$ as $r \rightarrow r_{+}$.
 This finding may have some implications for  the gravitational wave emission measured by an observer far away from the black holes \cite{LISAConsortiumWaveformWorkingGroup_2023}  as in the study of another interesting trajectories of homoclinic orbits in \cite{Levin_2009,Li_2023}.}

\section{Plunging ORBITS IN BOUND MOTION} \label{secIV}

Another bound orbit, in which particles eventually fall into the black hole, is the motion with the parameters of C and D in Fig. \ref{Bound_Plunge}.
In this case, there are two real roots, being  $r_{m1}$ inside the inner horizon, $r_{m4}$ outside the outer horizon, and a pair of the complex-conjugated roots $r_{m2}=r_{m3}^*$.
Assuming that the particle starts from $r_i \le r_{m4}$,
it will either plunge directly into the black hole or travel toward the root $r_{m4}$, return back and plunge into the black hole,} in the absence of any other real root along its trajectory.
%
This section is devoted to finding the analytical solution for the orbit in this case of $r_{m2}={r}^*_{m3}$ and $r_{m4}>r_i>r_{+}>r_{-}>r_{m1}$.
%

\begin{figure}[h]
 \centering
 \includegraphics[scale=0.5]{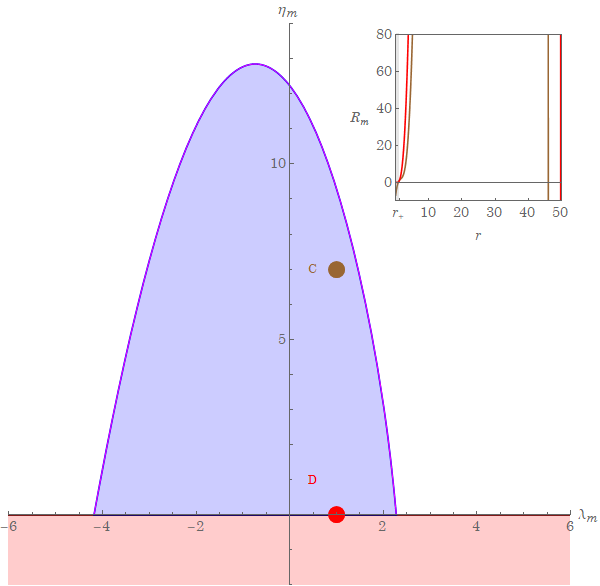}
 \caption{
{
The graphics shows the portion of parameter space bound by the double root solution, $r_{m2}=r_{m3}$.
The equation $R_m(r)$=0 with parameters in the blue zone have complex roots, $r_{m2}=r_{m3}^*$, so that, a particle in this region, say C or D, and starts from $r_i < r_4$ will plunge directly into the black hole horizon.
The inset shows the behavior of the radial potential $R_m(r)$ for the case of the parameters located in C and D.
}
\label{Bound_Plunge} }
\end{figure}

The solutions in the present cases are expressed in a similar form as in the previous section.
The integration of (\ref{r_theta}) is straightforward, but elliptical integrals and the Jacobi elliptic functions are involved for the representation of the solutions \cite{Abramowitz}.
We find after some algebra
\be\label{tauBmr}
\tau_{m}^{B}(r)=-\frac{1}{\sqrt{(1-\gamma_{m}^2)A_{m}B_{m}}}\left(F\left(\varphi(r)|k^{B}\right)-F\left(\varphi(r_{i})|k^{B}\right)\right)
\ee
where $F(\varphi|k)$ is the incomplete elliptic integral of the first kind. The two parameters of the elliptic integrals are
\be
\varphi(r)=\cos^{-1}\left(\frac{B_{m}(r_{m4}-r)-A_{m}(r-r_{m1})}{B_{m}(r_{m4}-r)+A_{m}(r-r_{m1})}\right)
\ee
and
\be
k^{B}=\frac{(r_{m4}-r_{m1})^2-(A_{m}-B_{m})^2}{4A_{m}B_{m}}\;, \label{k_B}
\ee
where we have used the short notations
\be
A_{m}=\sqrt{(r_{m4}-r_{m2})(r_{m4}-r_{m3})}\hspace{1mm},\hspace{1mm}B_{m}=\sqrt{(r_{m3}-r_{m1})(r_{m2}-r_{m1})}\,.\label{AB_B}
\ee
With the help of the Jacobian elliptic cosine function \cite{Abramowitz} one finds the inversion of (\ref{tauBmr}) as
\be
r^{B}(\tau_m)=\frac{(B_{m}r_{m4}+A_{m}r_{m1})-(B_{m}r_{m4}-A_{m}r_{m1}) {\rm cn}\left(X^{B}(\tau_m)\left|k^{B}\right)\right.}{(B_{m}+A_{m})-(B_{m}-A_{m}) {\rm cn}\left(X^{B}(\tau_m)\left|k^{B}\right)\right.}\;,\label{r_tau_m_u}
\ee
where
\begin{align}
&X^{B}(\tau_m)=\sqrt{\left(1-\gamma_m^2\right)A_{m}B_{m}}\tau_m-F\Bigg(\cos^{-1}\left(\frac{B_{m}(r_{m4}-r_i)-A_{m}(r_i-r_{m1})}{B_{m}(r_{m4}-r_i)+A_{m}(r_i-r_{m1})}\right)\left|k^{B}\Bigg)\right.\,.
\end{align}
%
Notice that $ A_m >B_m >0$, $0< k^B <1$, and for $r < r_{m4}$, $-1<\frac{B_{m}(r_{m4}-r_i)-A_{m}(r_i-r_{m1})}{B_{m}(r_{m4}-r_i)+A_{m}(r_i-r_{m1})}<1$. The Jacobian elliptic cosine function is  the real-valued function.

The solutions of the coordinates $\phi_m^B(\tau_m)$ and $t^B_m(\tau_m)$ involve the integrals $I_{m\phi}^{B}$ and $I_{mt}^{B}$ given in (\ref{Imphi_0}) and (\ref{Imt_0}).
%
In the present case, the integration of $I_1^B$, $I_2^B$, and $I_\pm^B$ is direct, but the results have cumbersome representations:
\begin{align}
&I_{\pm}^{B}(\tau_m)=\frac{1}{B_{m}\left(r_{m4}-r_{\pm}\right)+A_{m}\left(r_{\pm}-r_{m1}\right)}\left[\frac{B_{m}-A_{m}}{\sqrt{A_{m}B_{m}}}X^{B}(\tau_m)\right.\notag\\
&\quad\quad\quad\quad\quad\quad\quad\left.+\frac{2(r_{m4}-r_{m1})\sqrt{A_{m}B_{m}}}{B_{m}\left(r_{m4}-r_{\pm}\right)-A_{m}\left(r_{\pm}-r_{m1}\right)}R_{1}(\beta_{\pm}^{B};\Upsilon_{\tau_m}^{B}|k^{B}) \right]-\mathcal{I}_{\pm_i}^{B}\label{I pm tau m b}
\end{align}
\begin{align}
&I_{1}^{B}(\tau_m)=\left(\frac{B_{m}r_{m4}-A_{m}r_{m1}}{B_{m}-A_{m}}\right)\frac{X^{B}(\tau_m)}{\sqrt{A_{m}B_{m}}}+\frac{2(r_{m4}-r_{m1})\sqrt{A_{m}B_{m}}}{A_{m}^{2}-B_{m}^{2}}R_{1}(\beta^{B};\Upsilon_{\tau_m}^{B}|k^{B})-\mathcal{I}_{1_i}^{B}\label{I 1 tau m b}
\end{align}
\begin{align}
&I_{2}^{B}(\tau_m)=\left(\frac{B_{m}r_{m4}-A_{m}r_{m1}}{B_{m}-A_{m}}\right)^{2}\frac{X^{B}(\tau_m)}{\sqrt{A_{m}B_{m}}}\notag\\
&\quad\quad\quad\quad\quad\quad\left.+4\left(\frac{A_{m}r_{m1}-B_{m}r_{m4}}{A_{m}-B_{m}}\right)\frac{(r_{m4}-r_{m1})\sqrt{A_{m}B_{m}}}{A_{m}^{2}-B_{m}^{2}}R_{1}(\beta^{B};\Upsilon_{\tau_m}^{B}|k^{B})\right.\notag\\
&\quad\quad\quad\quad\quad\quad\quad+\sqrt{A_{m}B_{m}}\left(\frac{2(r_{m4}-r_{m1})\sqrt{A_{m}B_{m}}}{A_{m}^{2}-B_{m}^{2}}\right)^{2}R_{2}(\beta^{B};\Upsilon_{\tau_m}^{B}|k^{B})-\mathcal{I}_{2_i}^{B}\label{I 2 tau m b}
\end{align}
In the formulas above, the parameters of the functions $R_1$ and $R_2$ are related with the roots of $R_m(r)$ as follows
\begin{align}
&\hspace*{8mm}\beta_{\pm}^{B}=-\frac{B_{m}(r_{m4}-r_{\pm})+A_{m}(r_{\pm}-r_{m1})}{B_{m}(r_{m4}-r_{\pm})-A_{m}(r_{\pm}-r_{m1})}\;,\hspace*{4mm}\beta^{B}=\frac{A_{m}-B_{m}}{A_{m}+B_{m}}\,
\end{align}
\begin{align}\label{Upsilon_m_b}
&\Upsilon_{r}^{B}=\cos^{-1}\left(\frac{B_{m}(r_{m4}-r)-A_{m}(r-r_{m1})}{B_{m}(r_{m4}-r)+A_{m}(r-r_{m1})}\right),\hspace*{4mm}\Upsilon_{\tau_m}^{B}={\rm am}\left(X_{B}(\tau_m)\left|k_{B}\right)\right.
\end{align}
where $\rm am$ is the Jacobi amplitude function.
The quantities $\mathcal{I}_{\pm_i}^{B}$,  $\mathcal{I}_{1_i}^{B}$, and $\mathcal{I}_{2_i}^{B}$ are obtained  by evaluating ${I}_{\pm}^{B}$,  ${I}_{1}^{B}$, and ${I}_{2}^{B}$ at $r=r_i$ of the initial condition, ${{I}_{\pm}^{B}(0)={I}_{1}^{B}(0)={I}_{2}^{B}(0)=0}$.
Finally, $R_{1}$ and $R_{2}$ are the integral of Jacobian elliptic cosine function,
\be
R_{1}(\alpha;\phi|k)\equiv\int_{0}^{F(\phi|k)}\frac{du}{1+\alpha {\rm cn}(u|k)}=\frac{1}{1-\alpha^2}\left[\Pi\Bigg(\frac{\alpha^2}{\alpha^2-1};\phi\left|k\Bigg)\right.-\alpha f(p_\alpha,\phi,k)\right] \label{R1}
\ee
\begin{align}
&R_{2}(\alpha;\phi|k)\equiv\int_{0}^{F(\phi|k)}\frac{du}{[1+\alpha {\rm cn}(u|k)]^2}\notag\\
&\quad\quad=\frac{1}{\alpha^2-1}\left[F\left(\phi|k\right)-\frac{\alpha^2}{k+(1-k)\alpha^2}\left(E(\phi|k)-\frac{\alpha\sin(\phi)\sqrt{1-k\sin^2(\phi)}}{1+\alpha\cos(\phi)}\right)\right]\notag\\
&\quad\quad\quad+\frac{1}{k+(1-k)\alpha^2}\left(2k-\frac{\alpha^2}{\alpha^2-1}\right)R_{1}(\alpha;\phi|k) \label{R2}
\end{align}
in which
\begin{align}
&f(p_\alpha,\phi,k)=\frac{p_\alpha}{2}\ln\left(\frac{p_\alpha\sqrt{1-k\sin^2(\phi)}+\sin(\phi)}{p_\alpha\sqrt{1-k\sin^2(\phi)}-\sin(\phi)}\right)\, , \quad p_\alpha=\sqrt{\frac{\alpha^2-1}{k+(1-k)\alpha^2}}\, .
\end{align}
In particular, for $\alpha=\beta^B,\;\beta^B_\pm$ where $-1<\alpha<1$,  the solutions are the real-valued functions.

We then apply the exact solution obtained above to the parameters set C of Fig. \ref{Bound_Plunge}.
In this case, $\lambda_m=1$, $\eta_m=7$, and $\gamma_m=0.98$,
with the black hole parameters $a=0.7$ and $Q=0.7$.
Fig. \ref{OP_nonequa} shows that the particle stars from the initial position  $r_i=7.4M$, $\theta_i=\pi/2$, and $\phi_i=0$ and it falls almost directly into the black hole.

\begin{figure}[h]
 \centering
\includegraphics[scale=0.5,trim=40 60 100 100,clip]{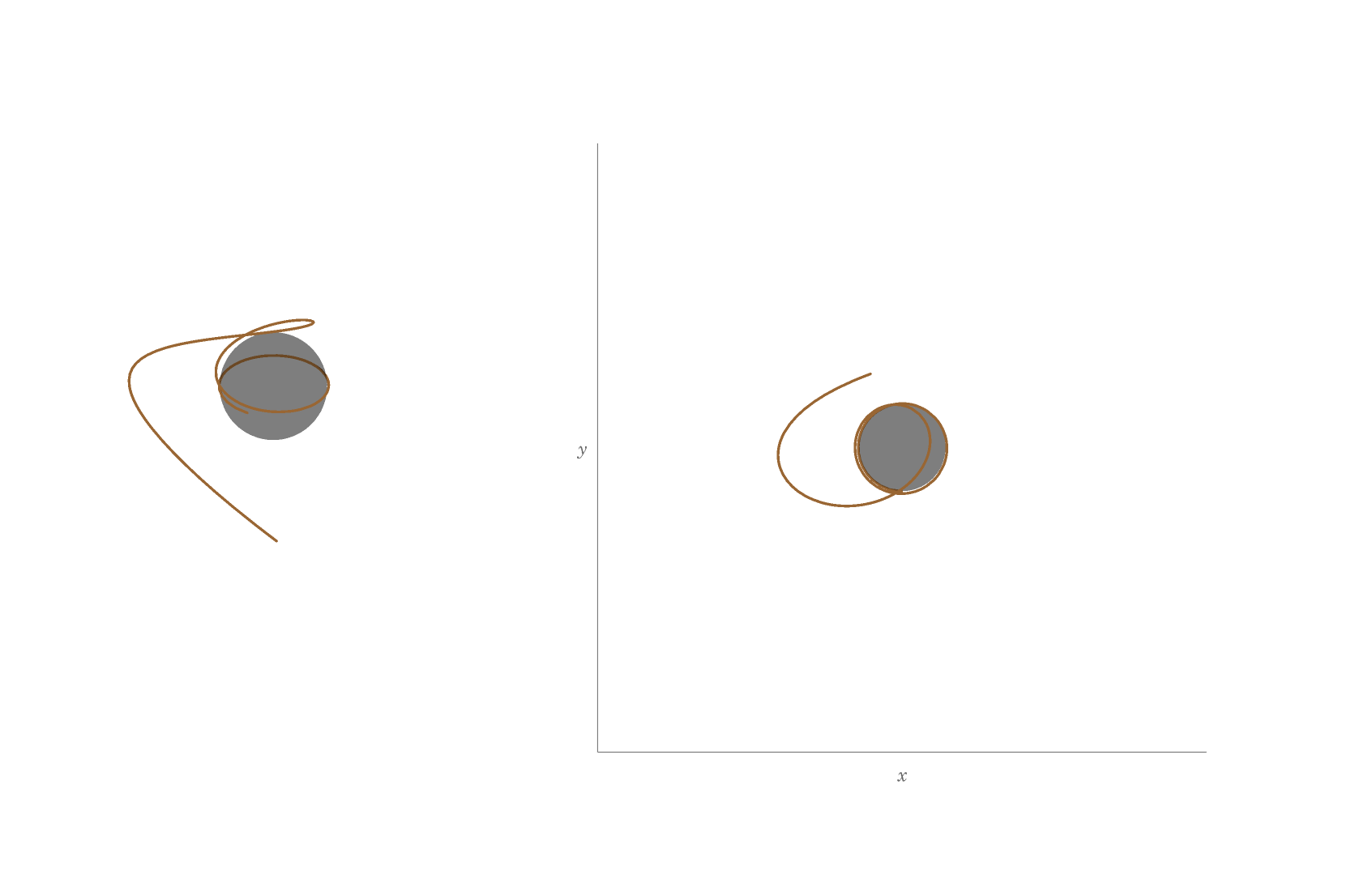}
 \caption{
{ Illustration of an orbit off the equatorial plane with the parameters of C in Fig. \ref{Bound_Plunge}. In this case the particle starts from $r_i < r_4$ and plunges directly into the black hole horizon. 
}
 \label{OP_nonequa}}
 \end{figure}

From the above general formulas one obtains the case of the equatorial motion, in which $\theta=\frac{\pi}{2}$ and $\eta_{m}\rightarrow0 $.
The bound plunge solution of the coordinates $\phi_m^{B}$ and  $t_m^{B}$ can be rewritten as the function of $r$ as follows
\begin{align}
&\phi_{m}^{B}\left(r\right)=I^B_{m\phi}\left(\tau_{m}\left(r\right)\right)+\lambda_{m}\tau^B_{m}\left(r\right)+\phi^B_{mi}\notag\\
&=\frac{\gamma_m}{\sqrt{1-\gamma_m^2}}\left[\frac{2Ma}{r_{+}-r_{-}}\left(\mathcal{J}_{m+}-\mathcal{J}_{m-}\right)-\frac{\lambda_{m}}{\gamma_{m}}f\left(r\right)\right] \; ,\label{phi m case 2 eq}
\end{align}
\begin{align}
&t_m^{B}\left(r\right)=I^B_{mt}\left(\tau_{m}\right)+t^B_{mi}\notag\\
&=\frac{\gamma_m}{\sqrt{1-\gamma_m^2}}\left\lbrace\frac{4M^2}{r_{+}-r_{-}}\left(\mathcal{T}_{m+}-\mathcal{T}_{m-}\right)+\frac{B_{m}r_{m4}-A_{m}r_{m1}}{B_{m}-A_{m}}\left(\frac{B_{m}r_{m4}-A_{m}r_{m1}}{B_{m}-A_{m}}+M\right)f\left(r\right)\right.\notag\\
&+\frac{2(r_{m4}-r_{m1})\sqrt{A_{m}B_{m}}}{A_{m}^2-B_{m}^2}\left[2\left(\frac{B_{m}r_{m4}-A_{m}r_{m1}}{B_{m}-A_{m}}\right)+M\right]R_{1}\left(\beta^{B};\varphi(r)|k^{B}\right)\notag\\
&\left.+4\sqrt{A_{m}B_{m}}\left[\frac{(r_{m4}-r_{m1})\sqrt{A_{m}B_{m}}}{A_{m}^2-B_{m}^2}\right]^2R_{2}\left(\beta^{B};\varphi(r)|k^{B}\right)+\left(4M^2-Q^2\right)f\left(r\right)\right\rbrace \;,\label{t m case 2 eq}
\end{align}
where
\begin{align}
\mathcal{T}_{m\pm}=\left(r_{\pm}-\frac{Q^2}{2M}\right)\mathcal{J}_{m\pm}\,,
\end{align}
\begin{align}
&\mathcal{J}_{m\pm}=\left(r_{\pm}-\frac{a\left(\frac{\lambda_{m}}{\gamma_{m}}\right)+Q^2}{2M}\right)\left\lbrace\frac{(B_{m}-A_{m})f\left(r\right)}{B_{m}(r_{m4}-r_{\pm})+A_{m}(r_{+}-r_{\pm})}\right.\notag\\
&\quad\quad\left.+\frac{2(r_{m4}-r_{m1})\sqrt{A_{m}B_{m}}}{\left[B_{m}(r_{m4}-r_{\pm})\right]^2-\left[A_{m}(r_{\pm}-r_{m1})\right]^2}R_{1}\left(\beta_{\pm}^{B};\varphi(r)|k^{B}\right)\right\rbrace\;,
\end{align}
\begin{align}
f\left(r\right)=\frac{1}{\sqrt{A_{m}B_{m}}}F\left(\varphi(r)|k^{B}\right)\,.
\end{align}
Fig. \ref{OP_equa} shows an exemplary orbit of this type using the these solutions.


\begin{figure}[h]
 \centering
 \includegraphics[scale=0.45]{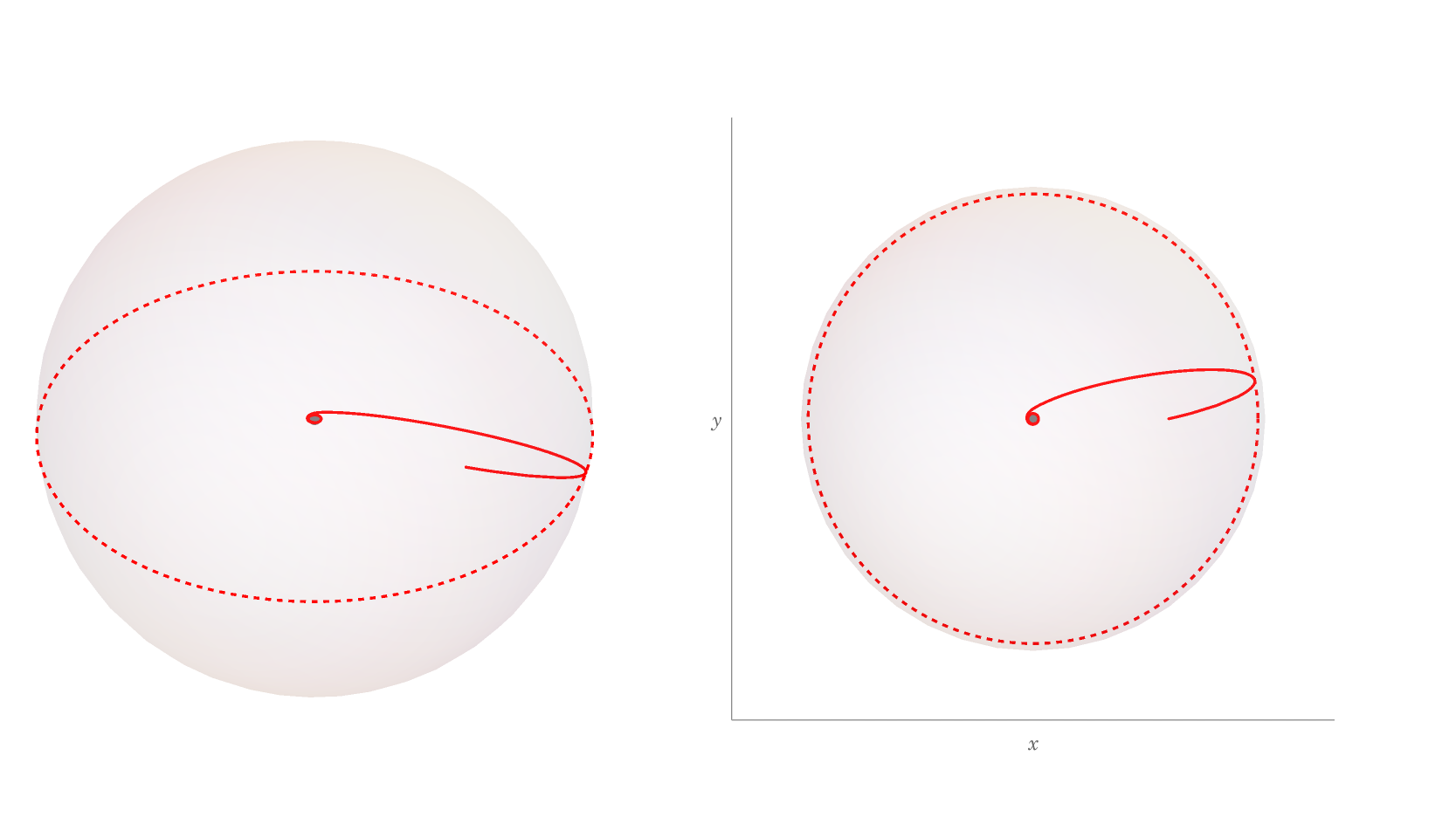}
 \caption{
{Illustration of an orbit on the equatorial plane with the parameters of D in Fig. \ref{Bound_Plunge}.
The particle initiates its journey at point $r_i$, moves outward, reaches the turning point at $r_{m4}$ , and then reverses its course, plunging back into the black hole.
}
  \label{OP_equa} }
 \end{figure}

The expression can also be converted into the solutions of the Kerr and  RN black holes by taking the respective $a\rightarrow0 $ and  $Q\rightarrow0 $ limit.
For the Kerr black hole, one can substitute straightforwardly $Q=0$ and the root $r_{m1}=0$ into the definition of $k^B$ and $B_m$ in (\ref{k_B}) and (\ref{AB_B}), as well as the solutions (\ref{phi m case 2 eq}) and (\ref{t m case 2 eq}).
Nevertheless, in the RN black hole the limits of $a\rightarrow 0$ but $r_{m1}\neq0$ give huge simplification. The formula (\ref{phi m case 2 eq}) becomes
\begin{align}
&\phi_{m}^{B}\left(r\right)=-\frac{\lambda_{m}}{\sqrt{(1-\gamma_{m}^2)A_{m}B_{m}}}F\Bigg(\cos^{-1}\left(\frac{B_{m}(r_{m4}-r)-A_{m}(r-r_{m1})}{B_{m}(r_{m4}-r)+A_{m}(r-r_{m1})}\right)\left|k^{B}\Bigg)\right.\,,\label{phi_m_B_KN}
\end{align}
whereas the solution of $t_m^{B}$ remains the same form as in (\ref{t m case 2 eq}) in the corresponding limits.
In the Schwarschild black hole where  $a,Q \rightarrow 0$, two event horizons, $r_{+}=2M$, $r_{-}=0$ giving $\mathcal{T}_{m-} \rightarrow 0$, together with $r_{m1}=0$, lead to the further simplification from (\ref{phi_m_B_KN})  and (\ref{t m case 2 eq}).

{One usefulness of the analytical formulas is to explore the changes of azimuthal angle $\phi$ as the motion crosses the horizon. In particular for the equatorial motion, as $r \rightarrow r_+$, the term of $\mathcal{J}_{m+}$ in (\ref{phi m case 2 eq}) due to the function of  $R_{1}(\alpha;\phi|k)$ in (\ref{R1}) gives the logarithmic divergence, namely $\phi_{m}^{B} \propto \ln (r-r_+)$. However, in the extremal case when $r_+=r_-=M$, the extra divergence occurs in the coefficient of $\mathcal{J}_{m+}-\mathcal{J}_{m+}$ so that the leading order divergence becomes  $\phi_{m}^{B} \propto \ln (r-M)$. In the case of $a=0$ for non spinning black hole, the coefficient of $\mathcal{J}_{m+}-\mathcal{J}_{m+}$ vanishes and the angle $\phi$ smoothly changes across the horizon. The corresponding plot is shown in Fig. \ref{phi_r_B}.}
\textcolor{blue}{}

\begin{figure}[h]
\centering
\includegraphics[width=0.8\columnwidth=1.0]{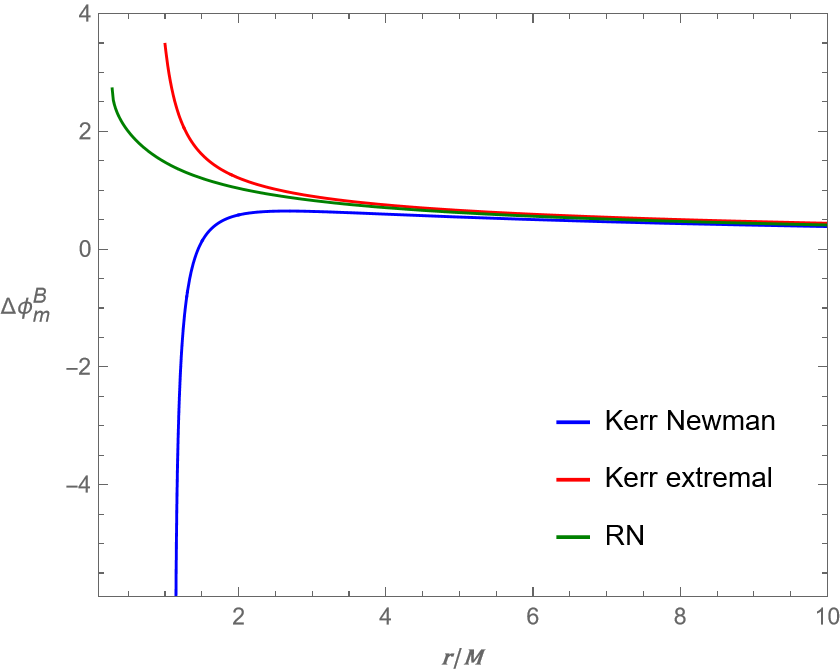}
\caption{
{
The plots show the $\Delta\phi_{m}^{B}$ as a function of $r$ for bound motion with the various black holes including Kerr-Newman ($Q=0.7 M, a=0.7 M, \lambda_m=M, \eta_m=0$), Kerr-extremal ($Q=0 M, a=M, \lambda_m=-M, \eta_m=0$) and RN ($Q=0.7 M, a=0 M, \lambda_m=M, \eta_m=0$).
In this plot, the red and blue curves illustrate the examples of direct and retrograde orbits, respectively.
}\label{phi_r_B}
}
 \end{figure}

\section{Plunging ORBITS IN UNBOUND MOTION}\label{secV}

For unbound motion ($\gamma_m >1$), the particle may start from the spatial infinity characterized by the constants of motion with the azimuthal angular  momentum $\lambda_m$, the energy $\gamma_m$, and the Carter constant $\eta_m$.
In this section we consider the parameters mainly in the E regime shown in Fig. \ref{Unbound_Plunge}, in which the roots of the radial potential have the properties, $r_{m3}^*={r}_{m4}$ and $r_i>r_{+}>r_{-}>r_{m2}>r_{m1}$.
This means that there is no turning point in the black hole exterior and the particle starting from the spatial infinity will plunge directly into the black hole.
%

\begin{figure}[h]
\centering
\includegraphics[scale=0.5]{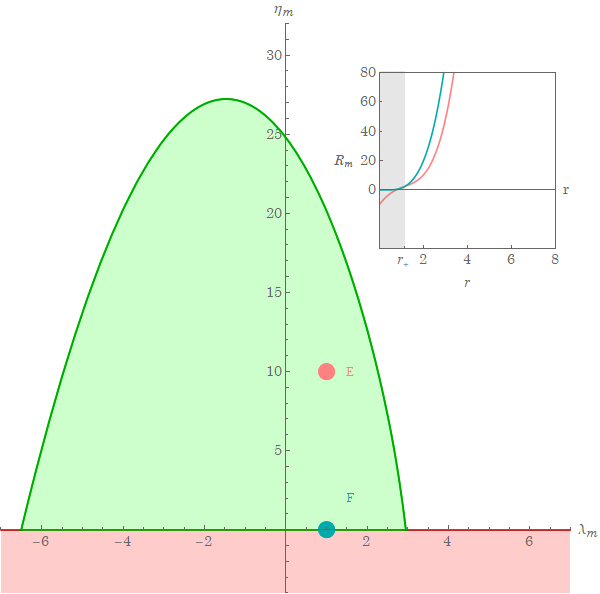}
\caption{
The graphics shows the portion of parameter space limited by the double root solution, $r_{m3}=r_{m4}$ and $r_{m1} < r_{m2} <r_-<r_+$.
For the region of the parameter space for E and F, the roots $r_{m3}$ and $r_{m4}$ are complex, $r_{m3}=r_{m4}^*$ and $r_{m1} < r_{m2} <r_-$.
The inset shows the details of the roots of illustrative cases E and F in the main figure . See the text for more discussion.\label{Unbound_Plunge}  }
\end{figure}

The main propose here is also to derive the exact solutions for the coordinates $r_m^U(\tau_m)$, $\theta_m^U(\tau_m)$, $\phi_m^U(\tau_m)$, and $t^U_m(\tau_m)$ (We have added the upper index $U$ for the unbound case).
While the procedure is identical to that of the previous two sections, special care is required due to differences in the properties of the roots.
The counterpart of Eq. (\ref{tauBmr}) becomes
\begin{align}
\tau^U_{m}=-\frac{1}{\sqrt{(\gamma_{m}^2-1)A_{m}^{U}B_{m}^{U}}}\left[F\left(\psi(r)|k^{U}\right)-F\left(\psi(r_{i})|k^{U}\right)\right] \,, \label{tauUm}
\end{align}
where
\begin{align}
&\psi(r)=\cos^{-1}\left(\frac{A_{m}^{U}(r-r_{m1})-B_{m}^{U}(r-r_{m2})}{A_{m}^{U}(r-r_{m1})+B_{m}^{U}(r-r_{m2})}\right)\,,
\end{align}
\begin{align}
&k^{U}=\frac{(A_{m}^{U}+B_{m}^{U})^2-(r_{m2}-r_{m1})^2}{4A_{m}^{U}B_{m}^{U}}\,,
\end{align}
and
\begin{align}
&A_{m}^{U}=\sqrt{(r_{m3}-r_{m2})(r_{m4}-r_{m2})} \;\;\;, \;\;\;
B_{m}^{U}=\sqrt{(r_{m3}-r_{m1})(r_{m4}-r_{m1})}\, .
\end{align}
Notice that $A^U_m$ and $B^U_m$ have different combinations of roots compared to  the bound case (\ref{AB_B}).
The evolution of the coordinate $r^U(\tau_m)$ is then
\begin{align}
&r^{U}(\tau_m)=\frac{(B_{m}^{U}r_{m2}-A_{m}^{U}r_{m1})+(B_{m}^{U}r_{m2}+A_{m}^{U}r_{m1}) {\rm { cn}}\left(X^{U}(\tau_m)\left|k^{U}\right)\right.}{(B_{m}^{U}-A_{m}^{U}){+}(B_{m}^{U}+A_{m}^{U}) {\rm { cn}}\left(X^{U}(\tau_m)\left|k^{U}\right)\right.}\,,\label{r_tau_m_u}
\end{align}
where
\begin{align}
&X^{U}(\tau_m)=\sqrt{\left(\gamma_m^2-1\right)A_{m}^{U}B_{m}^{U}}\tau_m-F\Bigg(\cos^{-1}\left(\frac{A_{m}^{U}(r_i-r_{m1})-B_{m}^{U}(r_i-r_{m2})}{A_{m}^{U}(r_i-r_{m1})+B_{m}^{U}(r_i-r_{m2})}\right)\left|k^{U}\Bigg)\right.\, .
\end{align}
%
Once again, the specified conditions—$B_m^{U} > A_m^{U} > 0$, $0 < k^U < 1$, and $r_{m1} < r_{m2} < r$, together with the inequality $-1 < \frac{A_{m}^{U}(r-r_{m1})-B_{m}^{U}(r-r_{m2})}{A_{m}^{U}(r-r_{m1})+B_{m}^{U}(r-r_{m2})} < 1$ ensure that the Jacobian elliptic cosine function in Eq. (\ref{r_tau_m_u}) remains a real-valued function.

The missing pieces for a complete description of the motions are the unbound version of equations (\ref{Imphi_0}) and (\ref{Imt_0}), in which the integral (\ref{I_n}) and (\ref{Ipm}) have been solved in Sec. \ref{secIV}. The results can be written as follows
%
%
\begin{align}
&I_{\pm}^{U}(\tau_m)=-\frac{1}{B_{m}^{U}\left(r_{\pm}-r_{m2}\right)+A_{m}^{U}\left(r_{\pm}-r_{m1}\right)}\left[\frac{B_{m}^{U}+A_{m}^{U}}{\sqrt{A^U_{m}B^U_{m}}}X^{U}(\tau_m)\right.\notag\\
&\quad\quad\quad\quad\quad\quad\quad\left.+\frac{2(r_{m2}-r_{m1})\sqrt{A_{m}^{U}B_{m}^{U}}}{B_{m}^{U}\left(r_{\pm}-r_{m2}\right)-A_{m}^{U}\left(r_{\pm}-r_{m1}\right)}R_{1}(\beta_{\pm}^{U};\Upsilon_{\tau_m}^{U}|k^{U}) \right]-\mathcal{I}_{\pm_i}^{U}\label{I pm tau m u}\;,
\end{align}
\begin{align}
&I_{1}^{U}(\tau_m)=\left(\frac{B_{m}^{U}r_{m2}+A_{m}^{U}r_{m1}}{B_{m}^{U}+A_{m}^{U}}\right)\frac{X^{U}(\tau_m)}{\sqrt{A_{m}^{U}B_{m}^{U}}}+\frac{2(r_{m2}-r_{m1})\sqrt{A_{m}^{U}B_{m}^{U}}}{(B_{m}^{U})^{2}-(A_{m}^{U})^{2}}R_{1}(\beta^{U};\Upsilon_{\tau_m}^{U}|k^{U})-\mathcal{I}_{1_i}^{U}\label{I 1 tau m u}\;,
\end{align}
\begin{align}
&I_{2}^{U}(\tau_m)=\left(\frac{B_{m}^{U}r_{m2}+A_{m}^{U}r_{m1}}{B_{m}^{U}+A_{m}^{U}}\right)^{2}\frac{X^{U}(\tau_m)}{\sqrt{A^U_{m}B^U_{m}}}\notag\\
&\quad\quad\quad\quad\quad\quad\left.+4\left(\frac{B_{m}^{U}r_{m2}+A_{m}^{U}r_{m1}}{B_{m}^{U}+A_{m}^{U}}\right)\frac{(r_{m2}-r_{m1})\sqrt{A_{m}^{U}B_{m}^{U}}}{(B_{m}^{U})^{2}-(A_{m}^{U})^{2}}R_{1}(\beta^{U};\Upsilon_{\tau_m}^{U}|k^{U})\right.\notag\\
&\quad\quad\quad\quad\quad\quad\quad\quad+\sqrt{A_{m}^{U}B_{m}^{U}}\left(\frac{2(r_{m2}-r_{m1})\sqrt{A_{m}^{U}B_{m}^{U}}}{(B_{m}^{U})^{2}-(A_{m}^{U})^{2}}\right)^{2}R_{2}(\beta^{U};\Upsilon_{\tau_m}^{U}|k^{U})-\mathcal{I}_{2_i}^{U}\;,\label{I 2 tau m u}
\end{align}
where the functions $R_1$ and $R_2$ have been defined in (\ref{R1}) and (\ref{R2}) and the unbound version of the parameters now read as
\begin{align}
&\hspace*{8mm}\beta_{\pm}^{U}=\frac{B_{m}^{U}(r_{\pm}-r_{m2})+A_{m}^{U}(r_{\pm}-r_{m1})}{B_{m}^{U}(r_{\pm}-r_{m2})-A_{m}^{U}(r_{\pm}-r_{m1})},\hspace*{4mm}\beta^{U}=\frac{B_{m}^{U}+A_{m}^{U}}{B_{m}^{U}-A_{m}^{U}}\,
\end{align}
\begin{align}\label{Upsilon_m_b}
&\Upsilon_{r}^{U}=\cos^{-1}\left(\frac{A_{m}^{U}(r-r_{m1})-B_{m}^{U}(r-r_{m2})}{A_{m}^{U}(r-r_{m1})+B_{m}^{U}(r-r_{m2})}\right),\hspace*{4mm}\Upsilon_{\tau_m}^{U}={\rm am}\left(X^{U}(\tau_m)\left|k^{U}\right)\right. \, .
\end{align}
As before, the initial conditions $\mathcal{I}_{\pm_i}^{U}$,  $\mathcal{I}_{1_i}^{U}$, $\mathcal{I}_{2_i}^{U}$ are obtained  by evaluating ${I}_{\pm}^{U}$,  ${I}_{1}^{U}$, ${I}_{2}^{U}$ at $r=r_i$ of the initial condition.
Also, for $\alpha$ in the functions (\ref{R1}) $R_1$ and $R_2$,  we have now $\alpha=\beta^U, \beta^U_\pm$, where $0<\alpha<1$, which ensure that the solutions are real-valued functions. Fig. \ref{U_O_3d} illustrates the orbits with the parameters of E in Fig. \ref{Unbound_Plunge}, where $\lambda_m=0$, $\eta_m=10$, and $\gamma_m=1.25$.

\begin{figure}[h]
\centering
\includegraphics[scale=0.5,trim=40 140 100 100,clip]{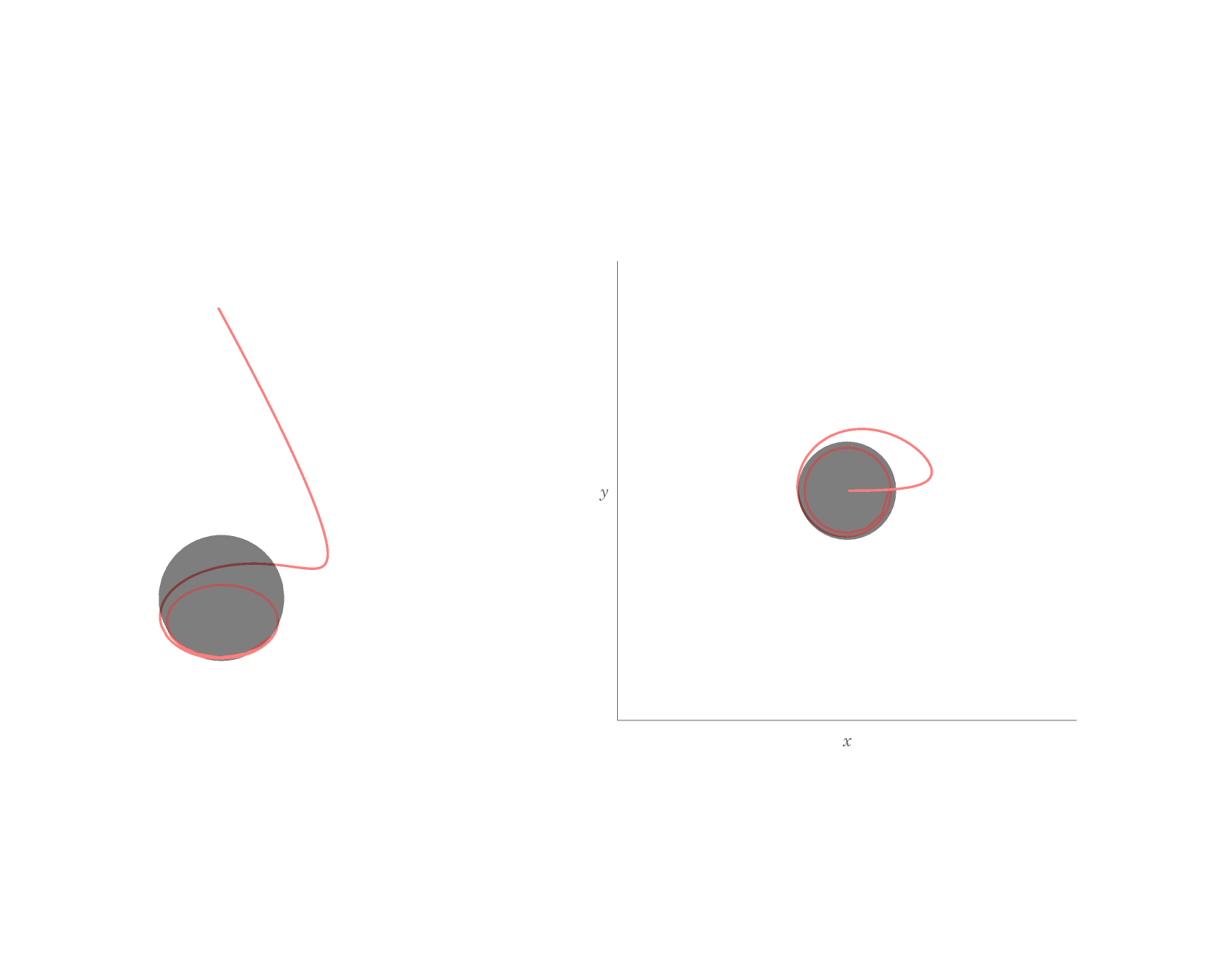}
\caption{
{
 Illustration of a general nonequatorial plunging orbit exemplifies with the parameters E in Fig. \ref{Unbound_Plunge}. In this case $\eta_m \neq 0$ and the particle starts from spatial infinity and infalls directly into the black horizon.} }
 \label{U_O_3d}
 \end{figure}

\begin{figure}[h]
 \centering
 \includegraphics[scale=0.5,trim=0 0 0 0,clip]{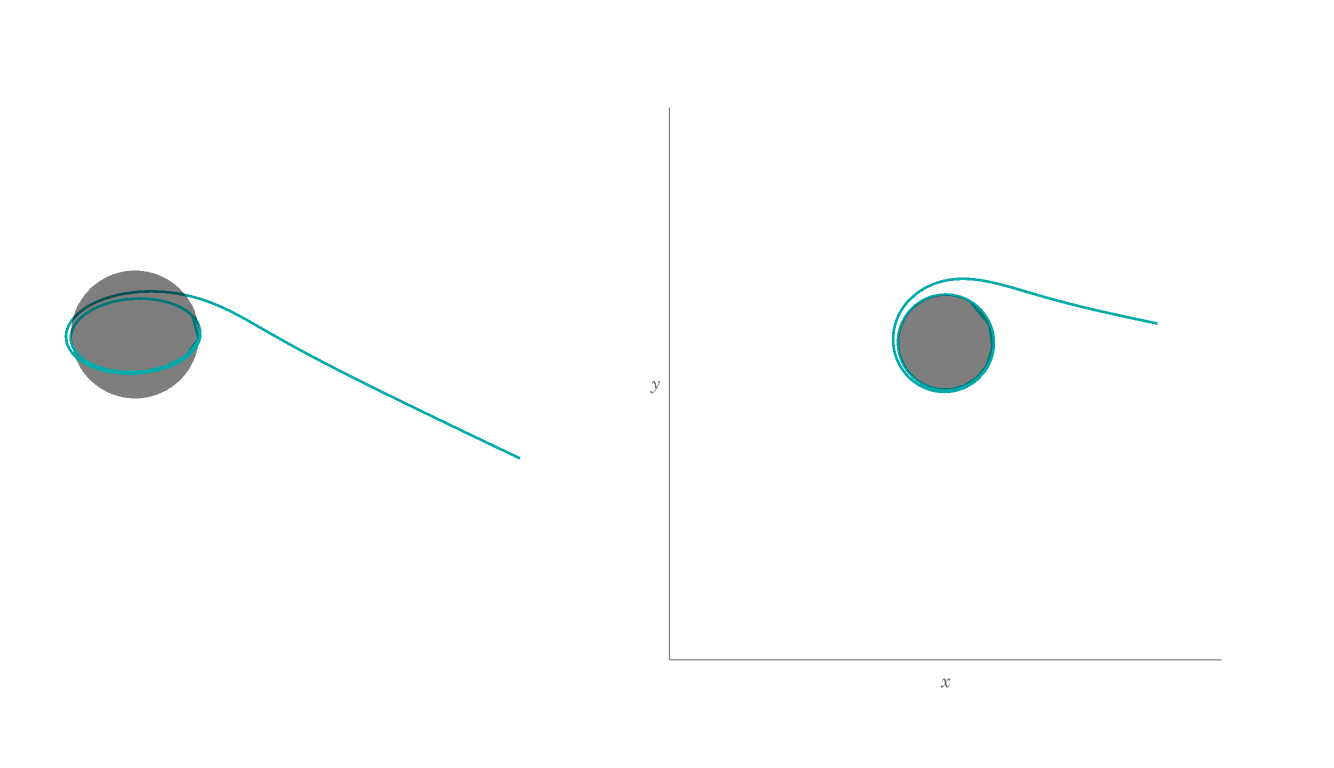}
 \caption{
{
 Illustration of an equatorial inspiral orbit with the parameters in the F region in Fig. \ref{Unbound_Plunge} for $\eta_m=0$ where the particle starts from spatial infinity and plunge directly into the horizon.} }
 \label{U_O_Equa}
 \end{figure}

Following the same steps of the previous sections, one can straightforwardly calculate the trajectories on the equatorial plane.
The solutions of $\phi_m^{U}$ and  $t_m^{U}$ as the function of $r$ can be written as follows
\begin{align}
&\phi_{m}^{U}\left(r\right)=I^U_{m\phi}\left(\tau_{m}\left(r\right)\right)+\lambda_{m}\tau_{m}\left(r\right)+\phi^U_{mi}\notag\\
&=\frac{\gamma_m}{\sqrt{\gamma_m^2-1}}\left[\frac{2Ma}{r_{+}-r_{-}}\left(\mathcal{K}_{m-}-\mathcal{K}_{m+}\right)-\frac{\lambda_{m}}{\gamma_{m}}g\left(r\right)\right]\, ,
\label{phi m case 3 eq}
\end{align}
\begin{align}
&t_m^{U}\left(r\right)=I^U_{mt}\left(\tau_{m}\right)+t^U_{mi}\notag\\
&=\frac{\gamma_m}{\sqrt{1-\gamma_m^2}}\left\lbrace\frac{4M^2}{r_{+}-r_{-}}\left(\mathcal{V}_{m-}-\mathcal{V}_{m+}\right)+\frac{B_{m}^{U}r_{m2}+A_{m}^{U}r_{m1}}{B_{m}^{U}+A_{m}^{U}}\left(\frac{B_{m}^{U}r_{m2}+A_{m}^{U}r_{m1}}{B_{m}^{U}+A_{m}^{U}}+M\right)g\left(r\right)\right.\notag\\
&+\frac{2(r_{m2}-r_{m1})\sqrt{A_{m}^{U}B_{m}^{U}}}{(B_{m}^{U})^2-(A_{m}^{U})^2}\left[2\left(\frac{B_{m}^{U}r_{m2}+A_{m}^{U}r_{m1}}{B^U_{m}+A^U_{m}}\right)+M\right]R_{1}\left(\beta^{U};\psi(r)|k^{U}\right)\notag\\
&\left.+4\sqrt{A_{m}^{U}B_{m}^{U}}\left[\frac{(r_{m2}-r_{m1})\sqrt{A_{m}^{U}B_{m}^{U}}}{(B_{m}^{U})^2-(A_{m}^{U})^2}\right]^{2}R_{2}\left(\beta^{U};\psi(r)|k^{U}\right)+\left(4M^2-Q^2\right)g\left(r\right)\right\rbrace\, \, , \label{t m case 3 eq}
\end{align}
where
\begin{align}
&\mathcal{V}_{m\pm}=\left(r_{\pm}-\frac{Q^2}{2M}\right)\mathcal{K}_{m\pm}\,,
\end{align}
\begin{align}
&\mathcal{K}_{m\pm}=\left(r_{\pm}-\frac{a\left(\frac{\lambda_{m}}{\gamma_{m}}\right)+Q^2}{2M}\right)\left\lbrace\frac{(B_{m}^{U}+A_{m}^{U})g\left(r\right)}{B_{m}^{U}(r_{\pm}-r_{m2})+A_{m}^{U}(r_{\pm}-r_{m1})}\right.\notag\\
&\quad\quad\left.+\frac{2(r_{m2}-r_{m1})\sqrt{A_{m}^{U}B_{m}^{U}}}{\left[B_{m}^{U}(r_{\pm}-r_{m2})\right]^2-\left[A_{m}^{U}(r_{\pm}-r_{m1})\right]^2}R_{1}\left(\beta_{\pm}^{U};\psi(r)|k^{U}\right)\right\rbrace\,,
\end{align}
\begin{align}
&g\left(r\right)=\frac{1}{\sqrt{A_{m}^{U}B_{m}^{U}}}F\left(\psi(r)|k^{U}\right) \, .
\end{align}
Fig. \ref{U_O_Equa} shows the orbit of the particle with the parameters in F in Fig. \ref{Unbound_Plunge}.

\begin{figure}[h]
\centering
\includegraphics[width=0.8\columnwidth=1.0]{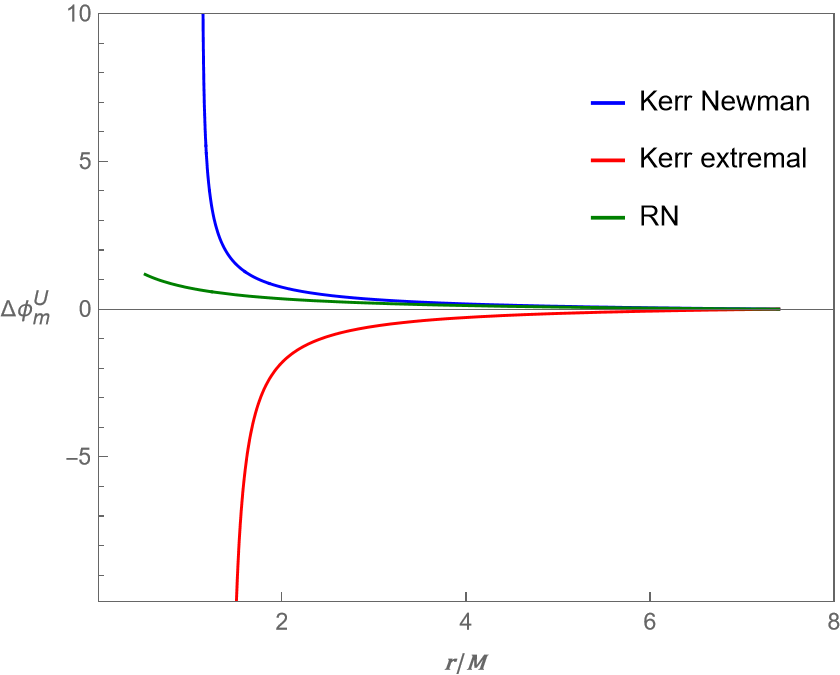}
\caption{
{
The plots show the $\Delta\phi_{m}^{U}$ as a function of $r$ for unbond infalling motion, with the various black hole including Kerr-Newman  ($Q=0.7 M, a=0.7 M, \lambda_m=M, \eta_m=0$), Kerr extremal ($Q=0 M, a=M, \lambda_m=-M, \eta_m=0$) and RN ($Q=0.7 M, a=0 M, \lambda_m=M, \eta_m=0$).
The blue and red curves show the examples of direct and retrograde orbits, respectively.
}
}\label{phi_r_U}
 \end{figure}

In the Kerr black hole, for $Q\rightarrow0 $, the solutions are given by taking $r_{m1}=0$ to  (\ref{phi m case 3 eq}) and (\ref{t m case 3 eq}). In the RN black hole for $a\rightarrow0 $ but $r_{m1}\neq0$, Eq. (\ref{phi m case 3 eq}) can be significantly simplified as
\begin{align}
&\phi_{m}^{U}\left(r\right)=-\frac{\lambda_{m}}{\sqrt{(\gamma_{m}^2-1)A_{m}^{U}B_{m}^{U}}}F\Bigg(\cos^{-1}\left(\frac{A_{m}^{U}(r-r_{m1})-B_{m}^{U}(r-r_{m2})}{A_{m}^{U}(r-r_{m1})+B_{m}^{U}(r-r_{m2})}\right)\left|k^{U}\Bigg)\right.\, .
\end{align}
In the Schwarschild black hole, $a, Q \rightarrow 0$ and the root $r_{m1}=0$ further simplify
the RN solution above.
Nevertheless, the solution of $t_m^{U}$ in various black holes remains the same form as (\ref{t m case 3 eq}) after taking the proper limits.

{The behavior of the azimuthal angle $\phi$ as the motion crosses the horizon shares the same features as in bound motion. In particular for the equatorial motion, as crossing the horizon, the term of $\mathcal{K}_{m_\pm}$ in (\ref{phi m case 3 eq}) due to the function of  $R_{1}(\alpha;\phi|k)$ in (\ref{R1}) gives the logarithmic divergence in general Kerr-Newman black holes. However, in the extremal case, the leading order divergence becomes  $\phi_{m}^{B} \propto \ln (r-M)$. In the case of $a=0$ for non spinning black hole, again the coefficient of $\mathcal{K}_{m+}-\mathcal{K}_{m+}$ vanishes and the angle $\phi$ smoothly changes across the horizon. The corresponding plot is shown in Fig. \ref{phi_r_U}.}


\section{Conclusions}\label{secVI}

In this paper, we analytically derive the solutions of infalling orbits in the context of general nonequatorial motion in the Kerr-Newman black holes, considering both bound and unbound motion.
These solutions can be written in terms of the elliptical integrals and the Jacobian elliptic functions of manifestly real functions of the Mino time.
Various limits have been taken to show the respective solutions in Kerr, Reissner-Nordstr$\ddot{o}$m, and Schwarzschild black holes.
In the case of the bound motion, we extend the study of \cite{mummery2022inspirals,Mummery_2023} on equatorial motion to consider that the particle  starts from $r \le r_{\rm ISSO}$ and then inspirals into the black hole on general nonequatorial motion.
In the limits of $Q \rightarrow 0$ and restricting on the equatorial plane, the obtained solutions reduces to the ones in \cite{mummery2022inspirals,Mummery_2023} .
We also consider the other types of the plunge motion,  with the values of
$\gamma_m$, $\lambda_m$, and $\eta_m$ shown in Fig.\ref{Bound_Plunge}.
In these cases, one has two real-valued roots of the radial potential, one inside the horizon, $r_{m1}$, and the other outside the horizon, $r_{m4}$.
Thus, the particle starts from $r \le r_{m4}$ and can either travel toward the $r_{m4}$, come back and then plunge directly into the black hole or travel directly into the black hole.
As for the unbound state, we showed the parameters $\gamma_m$, $\lambda_m$, and $\eta_m$  in Fig.\ref{Unbound_Plunge}.
Interestingly, while the two real-valued roots $r_{m2}$ and $r_{m1}$ are inside the event horizon, the other two roots are complex conjugate pair, $r_{m3}=r_{m4}^*$.
The particle starts from the spatial infinity and will plunge directly into the black holes.
%
{The analytical solutions allow us, in particular, to explore the behavior of the variation of azimuthal angle $\phi$ as the equatorial motion crosses the horizon.
In general Kerr-Newman black holes, the angle $\phi$ diverges as $\phi_{m} \propto \ln (r-r_+)$. However, in the extremal case, the leading order divergence becomes  $\phi_{m} \propto 1/(r-M)$.
In the case of non spinning black hole, the angle smoothly changes across the horizon.
{The dramatic difference in the behavior of the azimuthal angle $\phi$ across the horizon, which has been found in equatorial infalling orbits in Kerr black holes \cite{mummery2022inspirals, Mummery_2023}, is also seen in Kerr-Newman black holes in general nonequatorial orbits.  This may have some implications for the associated gravitational wave emission observed far away from the black holes.}

{These exact solutions of the spiral and plunge motions into the black hole  are also of astrophysical interest due to the fact that they have direct relevance to black hole accretion phenomena.  These explicit solutions may have applications to the numerical accretion  as well as extending current theories of black hole accretion \cite{FAB,PAG,REY,SCH}.}

\appendix

\section{The angular potential $\Theta(\theta)$ and the integrals $G_{m\theta}$, $G_{m\phi}$, and $G_{mt}$}\label{appA}

%
The detailed studies related to the $\Theta_m$ potential in the $\theta$ direction can be found in the papers \cite{Gralla_2020a,Wang_2022,Li_2023}. Here we summarize some of the relevant parts for the completeness of presentation.
The angular potential (\ref{Thetapotential}) for the particle can be rewritten in terms of $u=\cos^2 \theta$
and the equation of motion requires $\Theta_m\ge 0$, which restricts the parameter space of $\lambda_m$, $\eta_m$, and $\gamma_m$ (see Fig. 9 in \cite{Wang_2022}).
The roots of $\Theta_m(\theta)=0$ can be written as \cite{Gralla_2020a},
\begin{align}\label{u_m}
{u_{m,\pm}=\frac{\Delta_{m,\theta}\pm\nu_m \sqrt{\Delta_{m,\theta}^2+\frac{4\,{a}^2\, \eta_m}{\gamma_m^2-1}}}{2{a}^2}\,
\;\;
\Delta_{m \theta}={a}^2-\frac{\eta_m+\lambda_m^2}{\gamma_m^2-1}}
\end{align}
with 
{$\nu_m={\rm sign}(\gamma_m^2-1)$, which give the boundaries of the parameter space.}
%
For positive $\eta_m$ and nonzero $\lambda_m$ the particle starts off from the black hole exterior, ${1>u_+}>0$  is the only positive root, which in turn gives two roots at $\theta_{m+}=\cos^{-1}\left(-\sqrt{u_+}\right), \theta_{m-}=\cos^{-1}\left(\sqrt{u_+}\right)$.
The particle travels between the southern and northern hemispheres crossing the equator at $\theta=\frac{\pi}{2}$.

The solution of the coordinate $\theta_m(\tau_m)$ can be obtained by an
inversion of (\ref{r_theta})  \cite{Gralla_2020a,Wang_2022}
\be \label{theta_tau}
\theta(\tau_m)=\cos^{-1}\left(-\nu_{\theta_i}\sqrt{u_{m+}}{\rm sn}\left(\sqrt{-u_{m -}{a}^2\left(\gamma_m^2-1\right)}\left(\tau_m+\nu_{\theta_i}\mathcal{G}_{m\theta_i}\right)\left|\frac{u_{m+}}{u_{m-}}\right)\right.\right)\;,
\ee
where Mino time
\begin{align}
\tau_m=G_{m \theta}=p (\mathcal{G}_{m \theta_+}- \mathcal{G}_{m\theta_-}) + \nu_{\theta_i} \left[(-1)^p\mathcal{G}_{m\theta}-\mathcal{G}_{m\theta_i}\right] \label{tau_G_theta_m_a}
\end{align}
and sn denotes the Jacobi Elliptical sine function. In (\ref{tau_G_theta_m_a}) $p$ counts the times the trajectory passes through the turning points and $\nu_{\theta_i}={\rm sign}\left(\frac{d\theta_{i}}{d\tau'}\right)$.
The function $\mathcal{G}_{m \theta}$ is
\be \label{g_theta_m_a}
\mathcal{G}_{m\theta}=-\frac{1}{\sqrt{-u_{m-}\hat{a}^2\left(\gamma_m^2-1\right)}}F\left(\sin^{-1}\left(\frac{\cos\theta}{\sqrt{u_{m+}}}\right) \left|\frac{u_{m+}}{u_{m-}}\right)\right.\, .
\ee

The evolution of coordinates $\phi_m(\tau_m)$ and $t_m(\tau_m)$ in (\ref{phi}) and (\ref{t}) involves the integrals (\ref{Gmphi}) and (\ref{Gmt}), which can expressed as follows \cite{Wang_2022}
%
\begin{equation}
G_{m\phi}(\tau_m)=\frac{1}{\sqrt{-u_{m -}{a}^2\left(\gamma_m^2-1\right)}}\Pi\left(u_{m+};{\rm am}\left(\sqrt{-u_{m-}{a}^2\left(\gamma_m^2-1\right)}\left(\tau_m+\nu_{\theta_i}\mathcal{G}_{\theta_i}\right)\left|\frac{u_{m+}}{u_{m-}}\right)\right.\left|\frac{u_{m+}}{u_{m-}}\right)\right.-\nu_{\theta_i}\mathcal{G}_{m\phi_i}\,,\label{GmphiA}
\end{equation}
\begin{equation}
    \mathcal{G}_{\phi_i}=-\frac{1}{\sqrt{-u_{m-}{a}^2\left(\gamma_m^2-1\right)}}\Pi\left(u_{m+};\sin^{-1}\left(\frac{\cos\theta_i}{\sqrt{u_{m+}}}\right)\left|\frac{u_{m+}}{u_{m-}}\right)\right. \;,
\end{equation}
\begin{equation}
   G_{m t}(\tau_m)=-\frac{2u_{m+}}{\sqrt{-u_{m-}{a}^2\left(\gamma_m^2-1\right)}}E'\left({\rm am}\left(\sqrt{-u_{m-}{a}^2\left(\gamma_m^2-1\right)}\left(\tau_m+\nu_{\theta_i}\mathcal{G}_{m\theta_i}\right)\left|\frac{u_{m+}}{u_{m-}}\right)\right.\left|\frac{u_{m+}}{u_{m-}}\right)\right.-\nu_{\theta_i}\mathcal{G}_{m t_i}\,,\label{GmtA}
\end{equation}
\begin{equation}
   \mathcal{G}_{m t_i}=\frac{2u_{+}}{\sqrt{-u_{-}\hat{a}^2\left(\gamma_m^2-1\right)}}E'\left(\sin^{-1}\left(\frac
{\cos\theta_i}{\sqrt{u_{+}}}\right)\left|\frac{u_+}{u_-}\right)\right. \;.
\end{equation}
In the formulas above  $E$ and $\Pi$ are the incomplete elliptic integral of the second and third kinds, respectively.
%
Also the prime denotes the derivative with respect to the second argument,
\begin{align}
E'\left(\varphi\left|k\right)\right.=\partial_k E\left(\varphi\left|k\right)\right.=\frac{E\left(\varphi\left|k\right)\right.-F\left(\varphi\left|k\right)\right.}{2 k} \,.
\end{align}

\section{The radial potential $R_m(r)$ and its roots}\label{appB}

As for the radial potential (\ref{Rpotential}), it is a quartic polynomial. We then rewrite
$R_m({r})$ as follows
%
\begin{align}\label{R_m}
R_m({r})=S_m {r}^4+T_m {r}^3+U_m {r}^2+V_m {r}+W_m\,,
\end{align}
where the coefficients functions are given in terms constants of motion and parameters of the black hole as
\be
S_m=\gamma_m^2-1,
\ee
\be
T_m=2M,
\ee
\be
U_m={a}^2\left(\gamma_m^2-1\right)-{Q}^2-\eta_m-\lambda_m^2,
\ee
\be
V_m=2M\Bigl[\left({a}\gamma_m-\lambda_m\right)^2+\eta_m\Bigr],
\ee
\be
W_m=- {a}^2\eta_m-{Q}^2\Bigl[\left({a}\gamma_m-\lambda_m\right)^2+\eta_m\Bigr]\,.
\ee
Furthermore, it is useful to represent the radial potential using its roots, namely
%
\be
R_m({r})=\left(\gamma_m^2-1\right)({r}-r_{m 1})({r}-r_{m 2}) ({r}-r_{m 3}) ({r}-r_{m 4})\,.
\ee
The different dynamical behaviors of the system are characterized by the positions of these roots. See figures (\ref{Rm_r_triple}), (\ref{Bound_Plunge}), (\ref{Unbound_Plunge}), and also References \cite{Gralla_2020a,Wang_2022}.
The representation of roots of a quartic equation are well known, but cumbersome. We will write them down for the sake of unifying notation and ensuring the completeness of the work
%
\be
r_{m 1}=-\frac{M}{2\left(\gamma_m^2-1\right)}-z_m-\sqrt{-\hspace*{1mm}\frac{{X}_m}{2}-
z_m^2+\frac{{Y}_m}{4z_m}}\,,
\ee
\be
r_{m 2}=-\frac{M}{2\left(\gamma_m^2-1\right)}-z_m+\sqrt{-\hspace*{1mm}\frac{{X}_m}{2}-z_m^2+\frac{{Y}_m}{4z_m}}\,,
\ee
\be
r_{m 3}=-\frac{M}{2\left(\gamma_m^2-1\right)}+z_m-\sqrt{-\hspace*{1mm}\frac{{X}_m}{2}-z_m^2-\frac{{Y}_m}{4z_m}}\,,
\ee
\be
r_{m 4}=-\frac{M}{2\left(\gamma_m^2-1\right)}+z_m+\sqrt{-\hspace*{1mm}\frac{{X}_m}{2}-z_m^2-\frac{{Y}_m}{4z_m}}\,,
\ee
where
\be
z_m=\sqrt{\frac{\Omega_{m +}+\Omega_{m -}-\frac{{X}_m}{3}}{2}}\; ,
\ee
and
\be
\Omega_{m \pm}=\sqrt[3]{-\hspace*{1mm}\frac{{\varkappa}_m}{2}\pm\sqrt{\left(\frac{{\varpi}_m}{3}\right)^3+\left(\frac{{\varkappa}_m}{2}\right)^2}}\,
\ee
with
\be
{\varpi}_m=-\hspace*{1mm}\frac{{X}_m^2}{12}-{Z}_m \, , \quad\quad
{\varkappa}_m=-\hspace*{1mm}\frac{{X}_m}{3}\left[\left(\frac{{X}_m}{6}\right)^2-{Z}_m\right]-\hspace*{1mm}\frac{{Y}_m^2}{8}\,.
\ee
$X_m$, $Y_m$, and $Z_m$ are the short notation for
{\begin{align}
&{X}_m=\frac{8U_m S_m -3T_m^2}{8S_m^2}\,,\\
&{Y}_m=\frac{T_m^3-4U_m T_m S_m+8V_m S_m^2}{8S_m^3}\,,\\
&{Z}_m=\frac{-3T_m^4+256W_m S_m^3-64V_m T_m S_m^2+16U_m T_m^2S_m}{256S_m^4}\,.
\end{align}}

\begin{acknowledgments}
This work was supported in part by the National Science and Technology council (NSTC) of Taiwan, Republic of China.
\end{acknowledgments}

\end{document}